\documentclass[letterpaper,10pt,xcolor=dvipsnames,twoside]{article}

\usepackage{style}
\usepackage[english]{babel}
\usepackage[T1]{fontenc}
\usepackage[latin1]{inputenc}
\usepackage{mathrsfs}
\usepackage{amsmath,amssymb,amsfonts}
\usepackage{mathtools}						
\usepackage{graphicx}
\usepackage[colorlinks=true,linkcolor=blue, urlcolor=blue, anchorcolor=blue, citecolor=blue]{hyperref}
\usepackage{multirow}
\usepackage{multicol}
\usepackage{adjustbox}
\usepackage{cleveref}
\usepackage{hyperref}
\usepackage{dsfont}
\usepackage{xcolor}
\usepackage{float}

\usepackage{lettrine} 						
\usepackage{relsize}
\usepackage{chngcntr}
\usepackage{lineno} 						
\usepackage[hang,flushmargin]{footmisc}  	
\usepackage{setspace}

\crefname{equation}{Eq.}{Eqs.}
\crefname{figure}{Fig.}{Figs.}
\crefname{table}{Tab.}{Tabs.}
\Crefname{equation}{Equation}{Equations}
\Crefname{figure}{Figure}{Figures}
\Crefname{table}{Table}{Tables}

\DeclareMathAlphabet{\mathcal}{OMS}{cmsy}{m}{n}
\usepackage{mathrsfs}  
\usepackage{mathtools} 
\DeclareMathOperator{\atantwo}{atan2}

\usepackage{fancyhdr}
\let\OLDthebibliography\thebibliography
\renewcommand\thebibliography[1]{
	\OLDthebibliography{#1}
	\setlength{\parskip}{0pt}
	\setlength{\itemsep}{0pt plus 0.3ex}
}

\usepackage[font={sf,footnotesize}]{caption}
\usepackage[numbers,sort]{natbib}	
\newcommand*{\doi}[1]{\href{http://dx.doi.org/#1}{doi: #1}}

\numberwithin{equation}{section}
\counterwithout{equation}{section}

\newcommand{\um}{\text{\textmu m}}		
\newcommand{\I}{\operatorname{i}}		
\newcommand{\e}{\operatorname{e}}		

\newcommand{\ie}{i.\,e.\ }				
\newcommand{\ea}{\textit{et al.}}		

\newcommand{\dn}{\Delta n}				
\newcommand{\dmyelin}{t_{\text{sheath}}} 
\newcommand{\nmyelin}{n_{\text{m}}}		
\newcommand{\dk}{\Delta \kappa}			

\newcommand{\nex}{n_{\text{e}}}			
\newcommand{\nE}{n_{\text{E}}}			
\newcommand{\no}{n_{\text{o}}}			

\newcommand{\vE}{v_{\text{E}}}			
\newcommand{\vo}{v_{\text{o}}}			

\newcommand{\kex}{\kappa_{\text{e}}}	
\newcommand{\kE}{\kappa_{\text{E}}}		
\newcommand{\ko}{\kappa_{\text{o}}}		

\newcommand{\DS}{D_{\text{S}}}			
\newcommand{\DK}{D_{\text{K}}}			
\newcommand{\Dm}{|\mathscr{D}|}			

\newcommand{\phiP}{\varphi_{\text{P}}}  
\newcommand{\phiD}{\varphi_{\text{D}}}  
\newcommand{\deltaP}{\delta_{\text{P}}}  
\newcommand{\alphaP}{\alpha_{\text{P}}}  

\begin{document}


\begin{flushleft}
\Huge{ {\normalfont {\fontfamily{cmss}\selectfont
Diattenuation Imaging reveals different brain tissue properties
}}}\end{flushleft}

\vspace{0.2cm}

\begin{flushleft}{{\Large{{\normalfont{\fontfamily{cmss}\selectfont 
Miriam Menzel$^{1,2,\ast}$, Markus Axer$^{1}$, Katrin Amunts$^{1,3}$, Hans De Raedt$^{4}$ \\[8pt] \& Kristel Michielsen$^{5,2}$
}}}}}\end{flushleft}

\let\thefootnote\relax\footnotetext{
	{\fontfamily{cmss}\selectfont
		$^1$Institute of Neuroscience and Medicine (INM-1), Forschungszentrum Jülich GmbH, 52425 Jülich, Germany.
		$^2$Department of Physics, RWTH Aachen University, 52056 Aachen, Germany.
		$^3$Cécile and Oskar Vogt Institute for Brain Research, University Hospital Düsseldorf, University of Düsseldorf, 40204 Düsseldorf, Germany.
		$^4$Zernike Institute for Advanced Materials, University of Groningen, 9747AG Groningen, the Netherlands.
		$^5$Jülich Supercomputing Centre, Forschungszentrum Jülich GmbH, 52425 Jülich, Germany.\\
		$\ast$ Correspondence and requests for material should be addressed to M.M.\ (email: \href{mailto:m.menzel@fz-juelich.de}{m.menzel@fz-juelich.de})
}}

\vspace{1cm}


\begin{changemargin}{0cm}{4cm} 
\begin{spacing}{1.5}
{\large
{\fontfamily{cmss}\selectfont
\noindent 
When transmitting polarised light through histological brain sections, different types of diattenuation (polarisation-dependent attenuation of light) can be observed: In some brain regions, the light is minimally attenuated when it is polarised parallel to the nerve fibres (referred to as $D^+$), in others, it is maximally attenuated (referred to as $D^-$).
The underlying mechanisms of these effects and their relationship to tissue properties were so far unknown.
Here, we demonstrate in experimental studies that diattenuation of both types $D^+$ and $D^-$ can be observed in brain tissue samples from different species (rodent, monkey, and human) and that the strength and type of diattenuation depend on the nerve fibre orientations.
By combining finite-difference time-domain simulations and analytical modelling, we explain the observed diattenuation effects and show that they are caused both by anisotropic absorption (dichroism) and by anisotropic light scattering.
Our studies demonstrate that the diattenuation signal depends not only on the nerve fibre orientations but also on other brain tissue properties like tissue homogeneity, fibre size, and myelin sheath thickness.
This allows to use the diattenuation signal to distinguish between brain regions with different tissue properties and establishes Diattenuation Imaging as a valuable imaging technique.
\\[15pt]
}}
\end{spacing}
\end{changemargin}


\newpage
\begin{multicols}{2}
\small{

\lettrine{\textbf{A}}{} detailed knowledge of the brain's nerve fibre architecture and tissue composition is essential to understand its structure and function. Several neuroinflammatory and \mbox{-degenerative} disorders, for example multiple sclerosis \cite{lubetzki2014}, multiple system atrophy \cite{wenning2008,minnerop2010,ferrer2018}, and leukodystrophies \cite{knaap2017}, affect the integrity of neurons, axons, and dendrites. Therefore, it is important to study not only the long- and short-range connections between brain regions, but also the microstructure including axons and their myelin sheaths. 
In this paper, we show that the recently introduced technique \textit{Diattenuation Imaging} (DI), which compares measurements of nerve fibre orientations to \textit{diattenuation} (polarization-dependent attenuation), has the potential to provide new information about microstructural tissue properties.

The myelin content and connectivity patterns can be accessed in vivo by magnetic resonance imaging (MRI) techniques (T1, T2 \cite{glasser2011} and diffusion-weighting \cite{mori2006,tuch2003}) in the brains of healthy subjects and patients with a spatial resolution in the millimetre range \cite{chang2015}.
To study the nerve fibre architecture at microscopic resolution as prerequisite for a better interpretation and validation of MRI data, \textit{Three-dimensional Polarised Light Imaging} (3D-PLI) has been employed \cite{zeineh2016,henssen2018,caspers2017}.
In contrast to other microscopy techniques which are limited to smaller tissue samples, 3D-PLI allows to resolve three-dimensional nerve fibre pathways of unstained whole-brain sections at microscopic resolution:
The spatial orientations of the nerve fibres are derived by measuring the birefringence of the histological brain sections with a polarimeter \cite{MAxer2011_1,MAxer2011_2}. The birefringence is caused by highly ordered arrangements of nerve fibres as well as by the regular molecular structure of the myelin sheath \cite{schmitt1939,koike-tani2013,menzel2015} which surrounds most of the axons in the white matter \cite{martenson}, leading to negative birefringence with respect to the fibre direction \cite{menzel2015}. In the following, the term \textit{nerve fibre} will restrictively be used for myelinated axons.

The same anisotropy that causes birefringence (anisotropic refraction) also leads to diattenuation (anisotropic attenuation) \cite{mehta2013,chenault1993}, which is caused by anisotropic absorption (\textit{dichroism}) as well as by anisotropic scattering of light \cite{ghosh2011,chipman}.
The intensity of polarised light that is transmitted through a diattenuating medium depends on the direction of polarisation relative to the orientation of the optic axis (symmetry axis) in the medium: the transmitted light intensity becomes maximal for light polarised in a particular direction and minimal for light polarised in the corresponding orthogonal direction. 

There exist several studies that investigate the diattenuation of different biological tissue types (biopsy tissue \cite{soni2013}, skin \cite{jiao2003,ghosh2011}, heart \cite{fan2013}, muscle \cite{park2004}, tendon \cite{park2004}, collagen \cite{swami2006}, eye \cite{westphal2016}, unmyelinated nerve fibres in the retina \cite{naoun2005,huang2006}). Diattenuation has also been used to distinguish between healthy and pathological tissue (tissue from eye diseases \cite{naoun2005,huang2006}, burned/injured tissue \cite{jiao2003}, cancerous tissue \cite{soni2013}) and to quantify different tissue properties (e.\,g.\ concentration of glucose \cite{westphal2016,ghosh2011}, thickness \cite{naoun2005}). It can therefore be assumed that diattenuation studies of brain tissue would also reveal valuable additional information, assisting in the future to identify pathological changes.

In our previous study (\textsc{Menzel} \ea\ (2017) \cite{menzel2017}), we have explored the diattenuation of brain tissue for the first time. We have introduced Diattenuation Imaging (DI) as add-on to 3D-PLI: a combined measurement of diattenuation and birefringence compares the diattenuation signals of the brain section to the nerve fibre orientations obtained from 3D-PLI measurements. The developed measurement protocol allows to determine the diattenuation of whole unstained brain sections even with a low signal-to-noise ratio. By performing DI measurements of sagittal rat brain sections, we have found that there exist two different types of diattenuation that are regionally specific: in some brain regions, the transmitted light intensity becomes \textit{maximal} when the light is polarised parallel to the nerve fibres (referred to as $D^+$), in other brain regions, it becomes \textit{minimal} (referred to as $D^-$). 
Why this effect occurs and how the type of diattenuation depends on the tissue composition were still open questions.

In the present study, we modelled the observed diattenuation effects and provide an explanation for our previous results.
We demonstrate in experimental studies that diattenuation of both types $D^+$ and $D^-$ can be observed in brain tissue samples from different species (rodent, monkey, and human). By combining finite-difference time-domain simulations and analytical modelling, we explain the observed diattenuation effects and show that they are caused both by dichroism and by anisotropic light scattering. Our studies reveal that the diattenuation signal depends not only on the nerve fibre orientations and embedding time of the sample but also on tissue homogeneity, fibre size, and myelin sheath thickness. 
Thus, by comparing the diattenuation signals to the nerve fibre orientations, DI measurements allow to distinguish brain regions with different tissue composition. This helps not only to improve the reconstructed nerve fibre architecture but also to identify changes in brain tissue that are not visible in standard 3D-PLI measurements, making Diattenuation Imaging a valuable imaging technique in normal and pathologically altered tissue.


\section*{Results}

First, we performed DI measurements on brain sections from different species to investigate how the observed diattenuation effects ($D^+$ and $D^-$) depend on the tissue structure. 
To better understand the observed diattenuation effects and to learn more about how the diattenuation depends on underlying tissue properties, we subsequently simulated the diattenuation signal for different nerve fibre configurations.


\subsection*{Experimental studies}

The DI measurements \cite{menzel2017} were performed on 60\,\um\ thick brain sections as shown in \cref{fig:DI} (see \hyperref[methods]{Methods} for more details):
First, a 3D-PLI measurement was performed with a polarimeter (see \cref{fig:DI}a) to derive the three-dimensional nerve fibre orientations in each image pixel, \ie the in-plane orientation (\textit{direction}) angle $\phiP$ and the out-of-plane orientation (\textit{inclination}) angle $\alphaP$, see \cref{fig:DI}b (middle). The direction angle corresponds to the phase $\phiP$ of the measured birefringence signal, see \cref{fig:DI}b (left), the inclination angle is related to the amplitude $|\sin\deltaP|$ of the birefringence signal via: $\deltaP = (2\pi/\lambda)\, d \cos^2\alphaP$, where $\deltaP$ is the phase shift induced by the birefringent brain section, $\lambda$ the wavelength, $d$ the thickness of the birefringent brain section, and $\alphaP$ the inclination angle of the nerve fibres. The diattenuation of the brain section was measured with the same polarimeter (using only a rotating polariser, see \cref{fig:DI}d).
The amplitude (\textit{strength of diattenuation}) $\Dm$ and phase $\phiD$ of the measured diattenuation signal (see \cref{fig:DI}b, right) were computed from the maximum and minimum transmitted light intensities ($I_{\text{max}}$ and $I_{\text{min}}$) via \cite{chipman, chenault1993}: 
\begin{align}
\Dm \equiv \frac{I_{\text{max}} - I_{\text{min}}}{I_{\text{max}} + I_{\text{min}}}, 
\quad \phiD \equiv \varphi (I = I_{\text{max}})
\begin{dcases}
D^+: \phiD \approx \phiP \\
D^-: \phiD \approx \phiP + 90^{\circ},
\end{dcases}
\label{eq:D}
\end{align} 
where $\phiD$ denotes the direction of polarisation for which the transmitted light intensity becomes maximal.
A pixel-wise comparison of $\phiP$ and $\phiD$ (see histogram in \cref{fig:DI}c) reveals that some brain regions show diattenuation of type $D^+$ (highlighted in green), \ie the transmitted light intensity becomes maximal when the light is polarised \textit{parallel} to the nerve fibre direction ($\phiD \approx \phiP$), while other brain regions show diattenuation of type $D^-$ (highlighted in magenta), \ie the transmitted light intensity becomes maximal when the light is polarised \textit{perpendicularly} to the nerve fibre direction ($\phiD \approx \phiP + 90^{\circ}$). The values $\{\phiP, \phiD, \Dm\}$ were used to generate coloured \textit{diattenuation images} (see middle image in \cref{fig:DI}c): all $\Dm$ values that belong to regions with $(\phiD - \phiP) \in [-20^{\circ},20^{\circ}]$ were colourised in green ($D^+$), regions with $(\phiD - \phiP) \in [-70^{\circ},110^{\circ}]$ were colourised in magenta ($D^-$). The angle ranges account for the uncertainties of $\phiD$ due to the non-ideal optical properties of the \textit{Large-Area Polarimeter (LAP)} \cite{menzel2017} that was used for most DI measurements (see \hyperref[methods]{Methods}).

}\end{multicols} 

\begin{figure}[H]
	\centering
	\fbox{\includegraphics[width=0.95\textwidth]{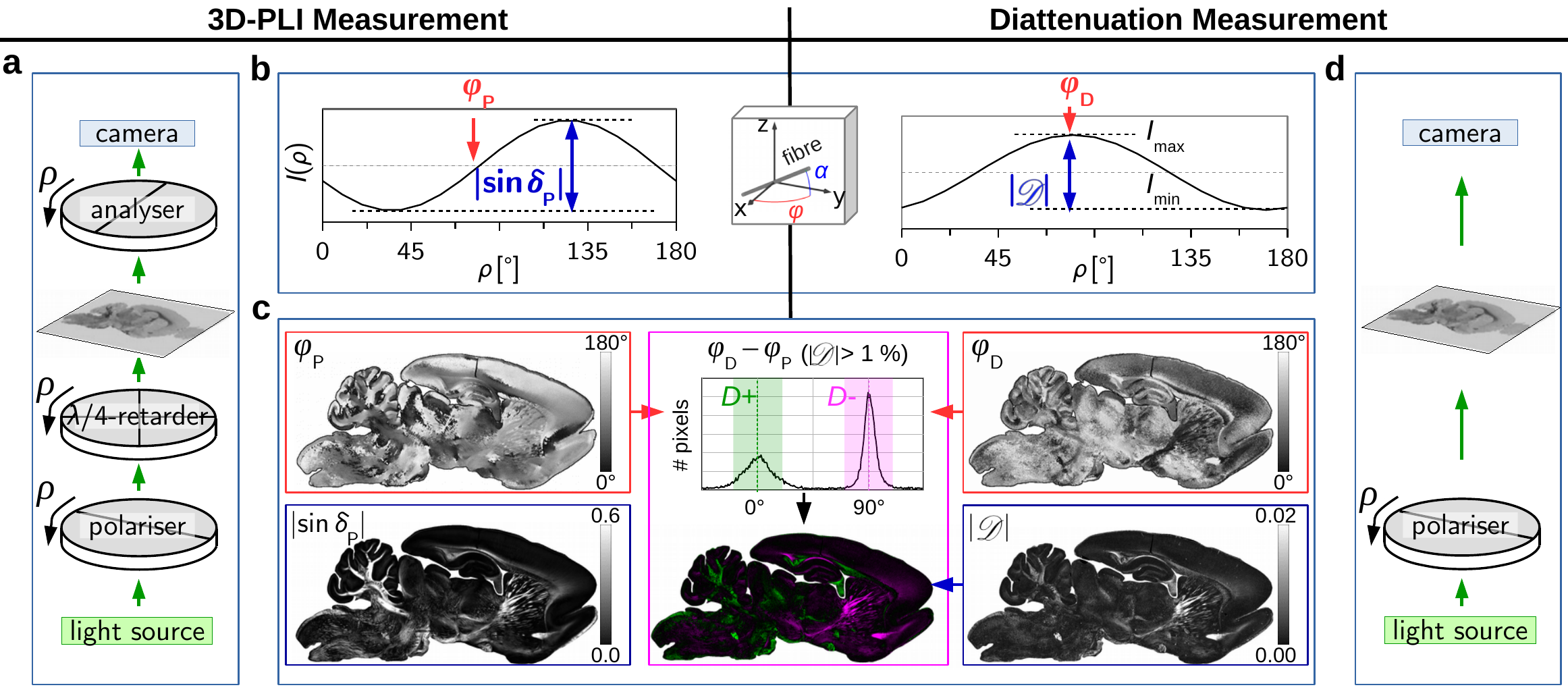}}
	\caption{\textbf{Diattenuation Imaging}: combined measurement of 3D-PLI (left) and diattenuation (right), shown exemplary for a sagittal rat brain section of 60\,\um\ thickness, measured with the LAP with an effective object-space resolution of 14\,\um\,/\,px (see \hyperref[methods]{Methods}). \textbf{a,d} Measurement set-up consisting of a pair of crossed linear polarisers and a quarter-wave retarder rotated by angles $\rho = \{0,10,\dots,170\}^{\circ}$. \textbf{b} 3D-PLI and diattenuation signals (transmitted light intensity $I(\rho)$). The phase and amplitude of the 3D-PLI signal ($\phiP$,$|\sin\deltaP|$) are directly related to the in-plane and out-of-plane orientation angles ($\varphi$,$\alpha$) of the nerve fibre, respectively. The phase and amplitude of the diattenuation signal ($\phiD$,$\Dm$) are derived from the maximum and minimum transmitted light intensities ($I_{\text{max}}$, $I_{\text{min}}$) using equation (\ref{eq:D}). \textbf{c} The coloured diattenuation image (middle image) is computed from the strength of the diattenuation signal $\Dm$, considering the phases $\{\phiP, \phiD\}$ of the 3D-PLI and diattenuation signals: all $\Dm$ values belonging to regions with $(\phiD - \phiP) \in [-20^{\circ},20^{\circ}]$ are colourised in green ($D^+$), regions with $(\phiD - \phiP) \in [-70^{\circ},110^{\circ}]$ are colourised in magenta ($D^-$), see peaks in histogram. The average transmitted light intensity $I_{\text{T}} = \overline{I(\rho)}$ of the brain section is shown in Supplementary \cref{fig:Diattenuation_vs_Time_Rat}a.}
	\label{fig:DI}
\end{figure}

\begin{figure}[!t]
	\centering
	\includegraphics[width=0.95\textwidth]{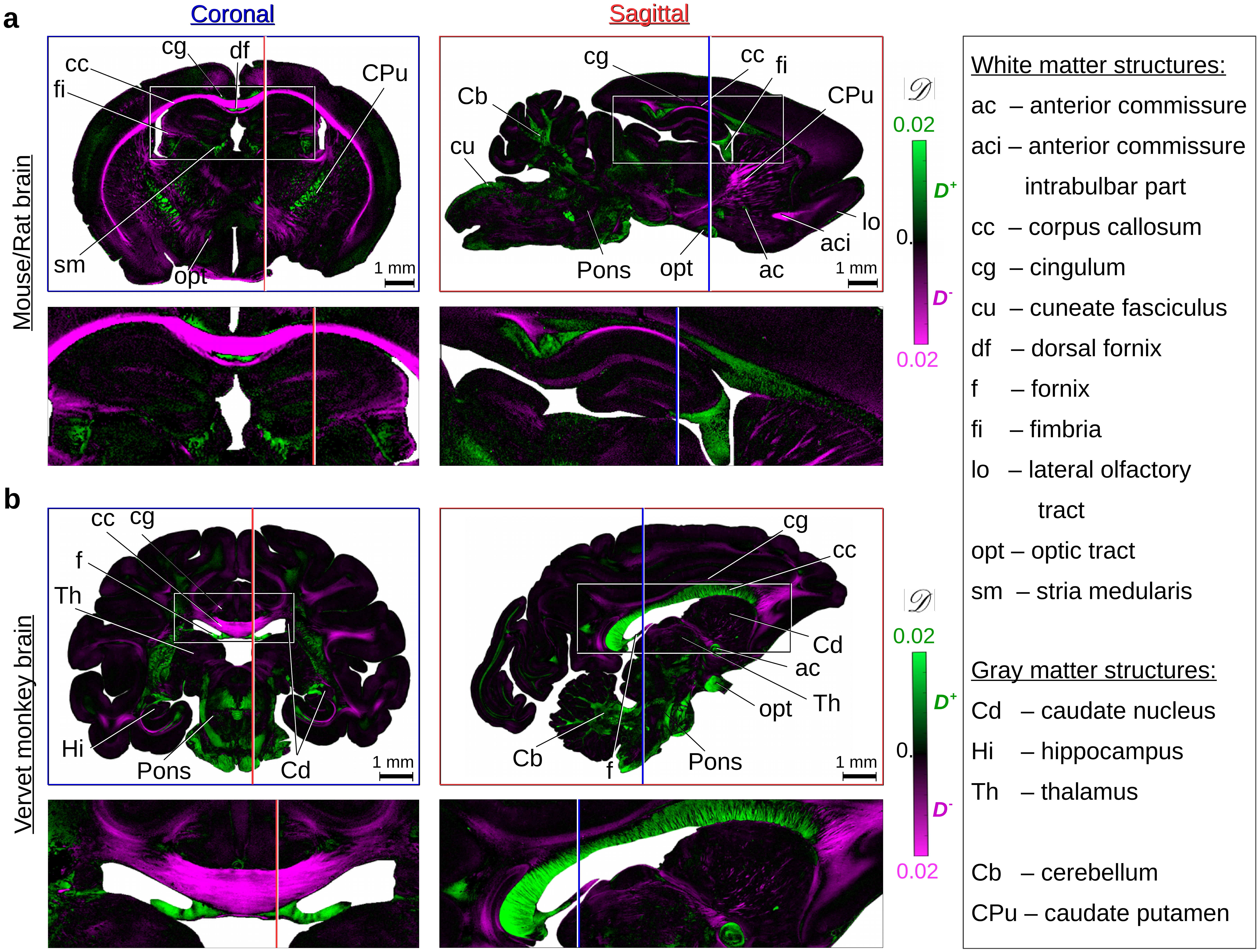}
	\caption{\textbf{Diattenuation images of coronal and sagittal brain sections} (60\,\um\ thickness). \textbf{a} Mouse brain (left) and rat brain (right) measured with an effective object-space resolution of 14\,\um\,/\,px. 
		\textbf{b} Vervet monkey brain measured with an effective object-space resolution of 27\,\um\,/\,px. 
		The coronal (sagittal) section planes are indicated by blue (red) lines in the respective other brain section for reference. 
		Regions surrounded by a white rectangle are shown as enlarged views.
		The strength of diattenuation $\Dm$ is shown in green (magenta) for regions with diattenuation of type $D^{+}$ ($D^{-}$), cf.\ \cref{fig:DI}c. The measurements were performed with the LAP one day after tissue embedding (see \hyperref[methods]{Methods}). Selected anatomical structures are labelled according to rat \cite{paxinos2007} and vervet \cite{woods2011} brain atlases. 
		Regions with $D^-$ mostly belong to regions with flat fibre structures (with respect to the section plane), regions with $D^+$ mostly belong to regions with steep fibre structures (cf.\ also \textsc{Menzel} \ea\ \cite{menzel2018-1} Fig.\ 1).}
	\label{fig:DiattenuationMaps_Coronal-Sagittal}
\end{figure}

\begin{multicols}{2} \small{ 


\paragraph{Diattenuation for different species and nerve fibre orientations.}
\Cref{fig:DiattenuationMaps_Coronal-Sagittal} shows the diattenuation images of coronal and sagittal brain sections for three different species: mouse (a), rat (a), and vervet monkey (b). 
The section planes are oriented perpendicularly to each other. For reference, the coronal (sagittal) section planes are indicated by blue (red) lines in the respective other brain section. 
The diattenuation measurements were performed with the LAP with an effective object-space resolution of 14\,\um\,/\,px (a) and 27\,\um\,/\,px (b).
The approximate orientation of the nerve fibres is known from brain atlases \cite{paxinos2007,woods2011} and 3D-PLI measurements \cite{zilles2015}.

Comparing the diattenuation images of coronal and sagittal brain sections to each other reveals that regions with diattenuation of type $D^-$ (magenta) mostly belong to regions with \textit{in-plane} fibre structures (coronal: cc, fi; sagittal: aci, cg, CPu), while regions with diattenuation of type $D^+$ (green) mostly belong to regions with \textit{out-of-plane} fibre structures (coronal: cg, df, sm, CPu; sagittal: cc, fi).
Fibre structures that show one type of diattenuation ($D^+$ or $D^-$) in one section plane (coronal or sagittal), are likely to show the other type of diattenuation in the other orthogonal section plane.

Measurements with a prototypic polarising microscope with 1.8\,\um\,/\,px yield similar $D^+$ and $D^-$ regions (see Supplementary \cref{fig:DiattenuationMaps_Taorad}), \ie the observed diattenuation effects do not depend on the optical resolution of the imaging system.

}\end{multicols} 

\begin{figure}[!t]
	\centering
	\includegraphics[width=0.6\textwidth]{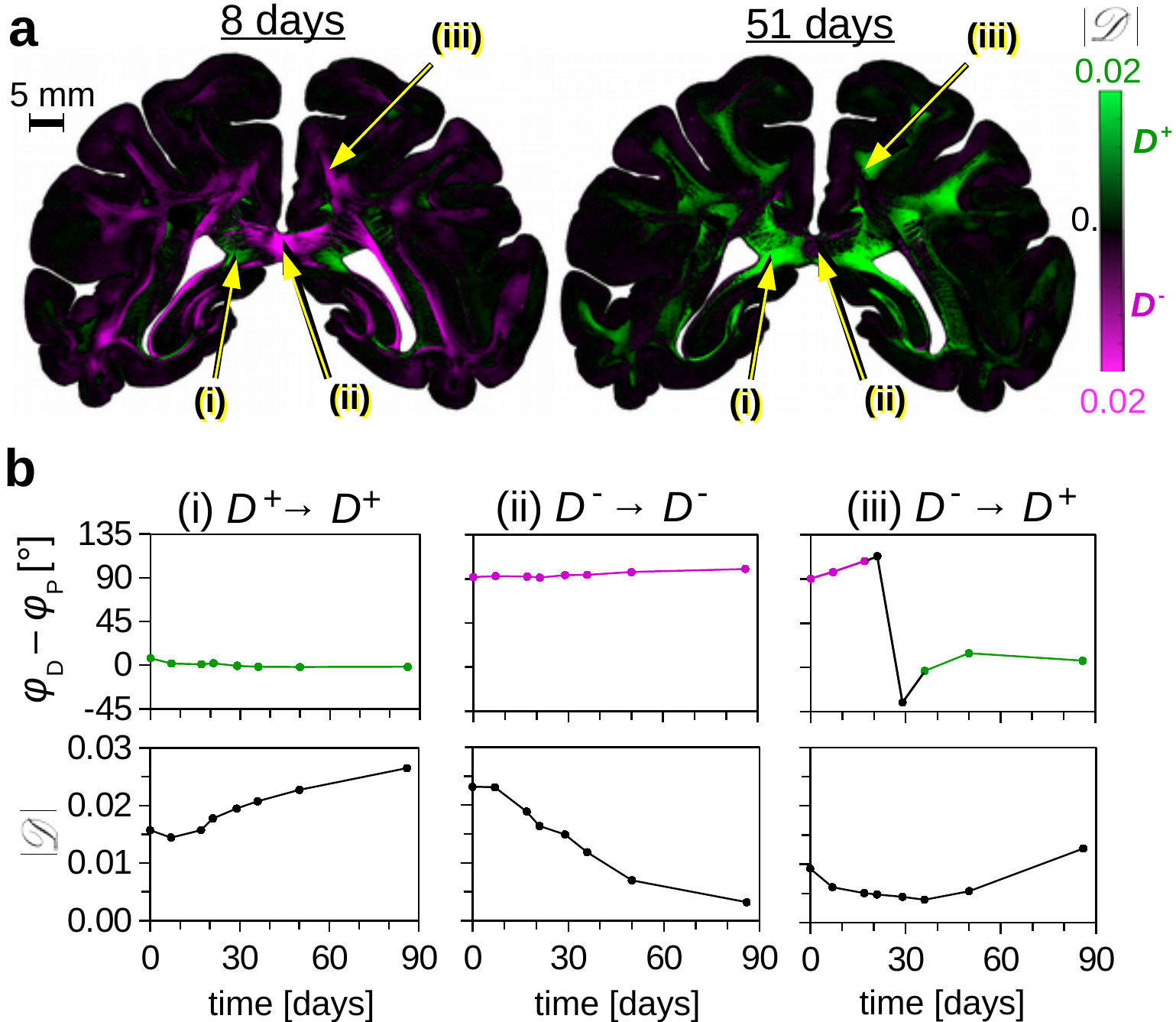}
	\caption{\textbf{Dependence of diattenuation on embedding time} (for a coronal vervet monkey brain section with 60\,\um\ thickness). \textbf{a} Diattenuation images measured 8 and 51 days after embedding the brain section in glycerine solution (effective object-space resolution: 43\,\um\,/\,px). Diattenuation values $\Dm$ that belong to regions with diattenuation of type $D^{+}$ ($\phiD - \phiP \in [-20^{\circ},20^{\circ}]$) are shown in green, diattenuation values that belong to regions with diattenuation of type $D^{-}$ ($\phiD - \phiP \in [70^{\circ},110^{\circ}]$) are shown in magenta, cf.\ \cref{fig:DI}c. The values $\{\Dm, \phiD, \phiP \}$ were determined from polarimetric measurements with the LAP (see \hyperref[methods]{Methods}). \textbf{b} Angle difference ($\phiD-\phiP$) and strength of the diattenuation signal $\Dm$ evaluated exemplary for three different regions, see yellow arrows in (a): a region that is already of type $D^+$ directly after tissue embedding (i), a region that is still of type $D^-$ after 51 days (ii), and a region that changes from type $D^-$ to $D^+$ over time (iii). In regions with $D^+$ ($D^-$), the strength of the diattenuation signal increases (decreases) with increasing time after tissue embedding.}
	\label{fig:Diattenuation_vs_Time_Vervet}
\end{figure}

\begin{multicols}{2} \small{ 
		

\paragraph{Dependence of diattenuation on embedding time.}
Prior to the measurements, the brain sections are embedded in glycerine solution (see \hyperref[methods]{Methods}).
To study how the diattenuation of a brain section changes with increasing time after tissue embedding, a coronal section of a vervet monkey brain was measured 1, 8, 18, 22, 30, 37, 51, and 87 days after tissue embedding with the LAP. \Cref{fig:Diattenuation_vs_Time_Vervet}a shows the diattenuation maps 8 days and 51 days after embedding. The size of the regions with diattenuation of type $D^+$ (green) increases with increasing time after embedding, while the size of the regions with diattenuation of type $D^-$ (magenta) decreases. The yellow arrows mark a region that is already of type $D^+$ directly after embedding (i), a region that is still of type $D^-$ after 51 days (ii), and a region that changes from type $D^-$ to $D^+$ (iii). \Cref{fig:Diattenuation_vs_Time_Vervet}b shows the corresponding values for ($\phiD - \phiP$) and $\Dm$ for all eight measurements.

In regions that show diattenuation of type $D^+$ ($D^-$), the strength of the diattenuation signal $\Dm$ increases (decreases).
The same pattern was observed in brain sections from other species including the human brain (see Supplementary \cref{fig:Diattenuation_vs_Time_Human}). 

With increasing embedding time, the diattenuation signal becomes more similar to the measured birefringence signal (cf.\ Supplementary \cref{fig:Diattenuation_vs_Time_Human}a,b). Brain sections with very long embedding times (several months) only show diattenuation of type $D^+$ (see Supplementary \cref{fig:Diattenuation_vs_Time_Rat}d).
		

\subsection*{Simulation Studies}

To model and better understand the observed effects, we simulated the measured diattenuation signals for different artificial nerve fibre configurations. 
As mentioned in the beginning, diattenuation can be caused both by anisotropic scattering ($\DS$) and by anisotropic absorption (dichroism $\DK$).

Diattenuation caused by anisotropic scattering was investigated by performing \textit{finite-difference time-domain (FDTD)} simulations and taking the optics of the polarimeter into account (see \hyperref[methods]{Methods}). The FDTD algorithm computes the propagation of the polarised light wave through the sample by approximating Maxwell's equations by finite differences \cite{taflove,menzel2016,menzel2018-1}. 

}\end{multicols} 

\begin{figure}[!t]
	\centering
	\includegraphics[width=\textwidth]{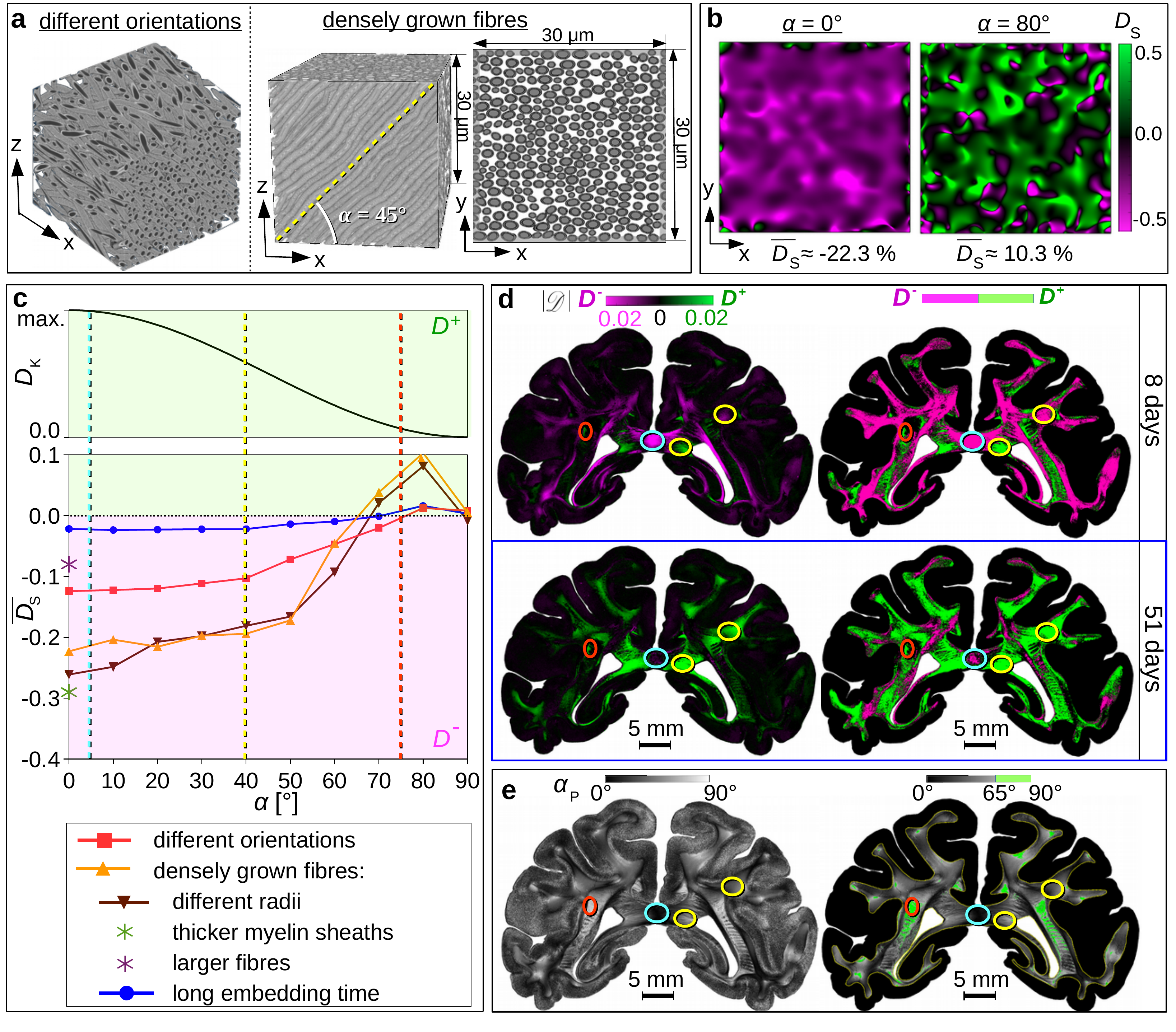}
	\caption{\textbf{Comparison of simulated and measured diattenuation effects.} \textbf{a} Fibre bundles used for FDTD simulations: a bundle with broad fibre orientation distribution (inclination $\alpha=0^{\circ}$) and a bundle of densely grown fibres ($\alpha=45^{\circ}$). \textbf{b} Simulated diattenuation images $\DS$ and mean values $\overline{\DS}$ for the bundle of densely grown fibres for $\alpha = 0^{\circ}$ and $80^{\circ}$. \textbf{c} Diattenuation caused by dichroism ($\DK$) and anisotropic scattering ($\DS$) plotted against $\alpha$. The curve for $\DK$ was computed analytically using $\DK = \tanh(0.05 \, \cos^2\alpha)$, see Supplementary \hyperref[note1]{Note1} for derivation. As the parameters depend on the exact fibre configuration and tissue composition, the curve is only qualitative and the maximum in arbitrary units. The curves for $\overline{\DS}$ were computed from FDTD simulations (see \hyperref[methods]{Methods}) for the bundle with broad fibre orientation distribution (red curve) and the bundle of densely grown fibres (orange curve) for fibres with radii $r \in [0.5, 0.8]\,\um$, myelin sheath thickness $\dmyelin = 0.35\,r$, and myelin refractive index $\nmyelin = 1.47$. The bundle of densely grown fibres was also simulated for a broader distribution of radii ($r \in [0.3, 1.0]\,\um$, brown curve) and for a sample with long embedding time ($\nmyelin = 1.39$, blue curve). For the horizontal case ($\alpha = 0^{\circ}$), the bundle was also simulated for thicker myelin sheaths ($\dmyelin = 0.6\,r$, green star) and larger fibres ($r \in [2.5, 4.0]\,\um$, magenta star). Positive (negative) diattenuation values which correspond to type $D^+$ ($D^-$) are shown on a green (magenta) background. \textbf{d} Diattenuation images of a coronal vervet brain section measured 8 and 51 days after tissue embedding (adapted from \cref{fig:Diattenuation_vs_Time_Vervet}a). Diattenuation values $\Dm$ that belong to regions with diattenuation of type $D^{+}$ ($D^-$) are shown in green (magenta), regions that cannot clearly be assigned are shown in black. The images on the left show the strength of the diattenuation signal $\Dm$ coloured according to the type of diattenuation. The images on the right separately show the $D^+$ and $D^-$ regions in the white matter. \textbf{e} Corresponding inclination angles $\alphaP$ obtained from a 3D-PLI measurement with tilting (see \hyperref[methods]{Methods}). The image on the right only shows the inclination angles in the white matter and regions with $\alphaP > 65^{\circ}$ are marked in green. The dashed vertical lines in (c) and the coloured circles in (d) and (e) highlight regions with flat, intermediate, and steep fibre inclinations: $\alpha = 5^{\circ}$ (cyan), $\alpha = 40^{\circ}$ (yellow), and $\alpha = 75^{\circ}$ (red). }
	\label{fig:Diattenuation_Sim_vs_Exp}
\end{figure}

\begin{multicols}{2} \small{ 
		
In the diattenuation measurement, the polariser is rotated by 18 discrete angles. To save computing time, the diattenuation signal was approximately computed from only two simulation runs: from the transmitted intensity of light polarised along the x-axis ($I_{\text{x}}$) and from the transmitted intensity of light polarised along the y-axis ($I_{\text{y}}$), where the x-axis is aligned with the symmetry axis of the sample projected onto the xy-plane:
\begin{align}
\DS \equiv \frac{I_{\text{x}} - I_{\text{y}}}{I_{\text{x}} + I_{\text{y}}}, \,\,\,
-1 \leq \DS \leq 1  \,\,
\begin{dcases}
\DS > 0 \,\, \Leftrightarrow \, D^+: \phiD \approx \phiP, \\
\DS < 0 \,\, \Leftrightarrow \, D^-: \phiD \approx \phiP + 90^{\circ}.
\end{dcases}
\label{eq:DS}
\end{align}
The magnitude of $\DS$ is related to the strength of the diattenuation ($\vert \DS \vert \approx \Dm$), the sign indicates the phase $\varphi_{\text{D}}$ (cf.\ equation (\ref{eq:D})):
Positive values ($\DS > 0 \Leftrightarrow I_{\text{x}} > I_{\text{y}}$) correspond to regions with $D^+$ and are shown in green (the transmitted light intensity becomes maximal when the light is polarised parallel to the fibre structure, \ie in the x-direction). Negative diattenuation values ($\DS < 0 \Leftrightarrow I_{\text{x}} < I_{\text{y}}$) correspond to regions with $D^-$ and are shown in magenta (the transmitted light intensity becomes maximal when the light is polarised perpendicularly to the fibre structure, \ie in the y-direction).

The diattenuation caused by dichroism was described by an effective analytical model (see upper panel in \cref{fig:Diattenuation_Sim_vs_Exp}c), assuming that birefringence and dichroism can be described by a complex retardance with shared principal axes (see Supplementary \hyperref[note1]{Note1} for derivation): $\DK \approx \tanh(-(2\pi/\lambda)\,d\,\Delta\kappa \cos^2\alpha)$, where $\lambda$ is the wavelength, $d$ the thickness of the dichroic brain section, $\alpha$ the inclination angle of the nerve fibres, and $\Delta\kappa$ the anisotropic absorption (dichroism) of the brain tissue. As the birefringence of the brain sections does not change much with increasing embedding time (see Supplementary \cref{fig:Diattenuation_vs_Time_Rat}b), the dichroism is also expected to be mostly time-independent. In contrast, the diattenuation caused by anisotropic scattering is expected to decrease with increasing embedding time because brain sections with long embedding times become transparent, \ie show less scattering (see Supplementary \cref{fig:Diattenuation_vs_Time_Rat}a). As brain sections with long embedding times only show diattenuation of type $D^+$, dichroism was assumed to cause positive diattenuation: $\DK > 0 \Leftrightarrow \Delta\kappa < 0$.


\paragraph{Simulated diattenuation for different fibre configurations.}
To analyse how the diattenuation of brain tissue depends on the out-of-plane orientation (inclination angle) of the enclosed nerve fibres, we simulated the diattenuation images for fibre bundles with different inclination angles ($\alpha = \{0^{\circ}, 10^{\circ}, \dots, 90^{\circ}\}$): a bundle with broad fibre orientation distribution and a bundle of densely grown fibres (see \cref{fig:Diattenuation_Sim_vs_Exp}a).
\Cref{fig:Diattenuation_Sim_vs_Exp}b shows the diattenuation images obtained from FDTD simulations for the bundle of densely grown fibres for $\alpha = 0^{\circ}$ and $80^{\circ}$. 
The diattenuation images for all inclination angles are shown in Supplementary \cref{fig:Sim_DiattenuationMaps}.
Regions with maximum and minimum diattenuation values are homogeneously distributed. 

}\end{multicols} 

\begin{figure}[H]
	\centering
	\includegraphics[width=0.8\textwidth]{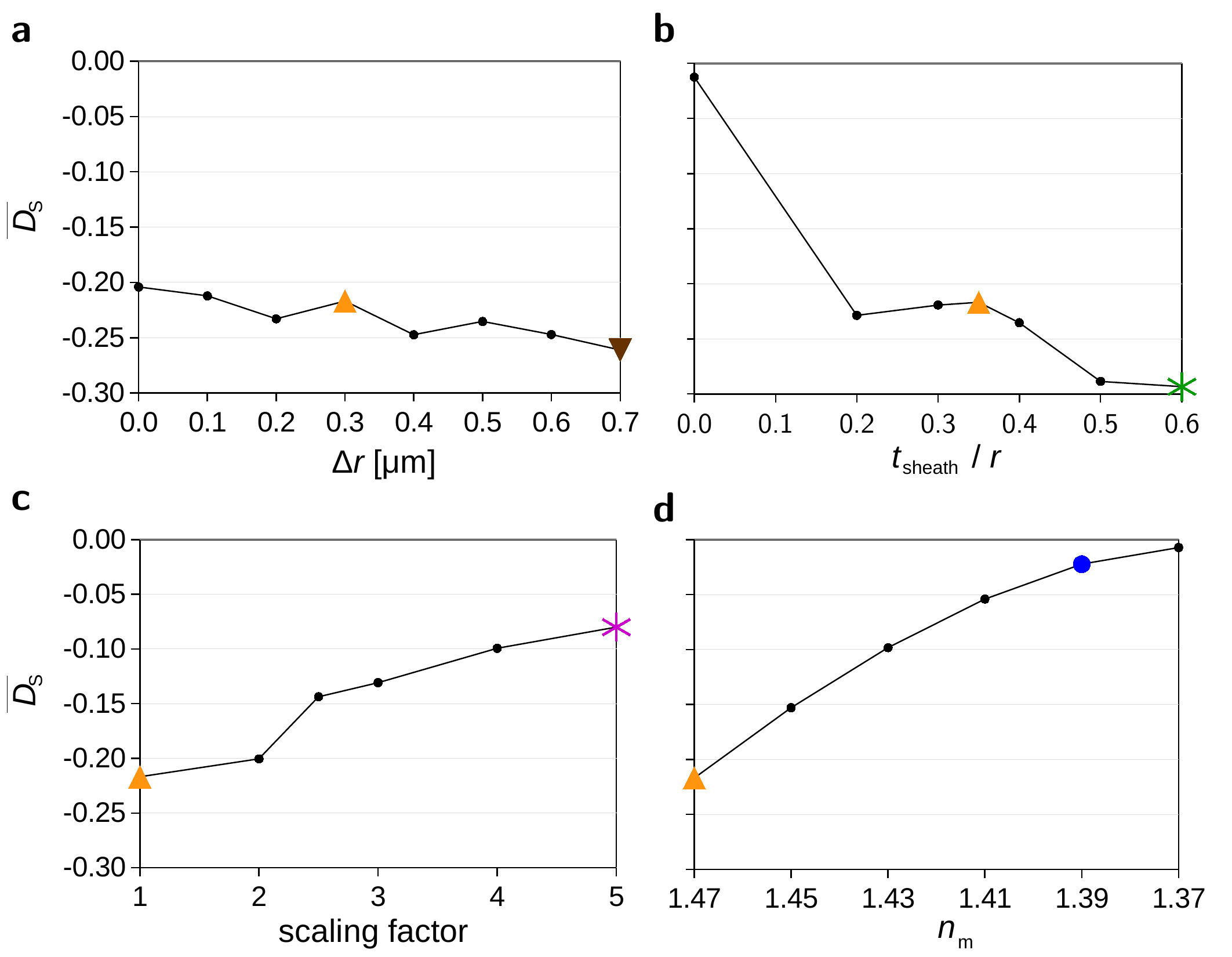}
	\caption{Mean diattenuation values $\overline{\DS}$ obtained from simulations of the horizontal bundle of densely grown fibres (cf.\ \cref{fig:Diattenuation_Sim_vs_Exp}a with $\alpha = 0^{\circ}$) for different fibre properties (see \hyperref[methods]{Methods}): \textbf{a} different fibre radius distributions $r \in [r_{\text{min}}, r_{\text{max}}]$ with $\Delta r \equiv r_{\text{max}} - r_{\text{min}}$ and $(r_{\text{max}} + r_{\text{min}})/2 = 0.65\,\um$, \textbf{b} different myelin sheath thicknesses $t_{\text{sheath}}$ (relative to the fibre radius $r$), \textbf{c} different scaling factors (fibre sizes), \textbf{d} different myelin refractive indices $n_{\text{m}}$. The values belonging to the graphs in \cref{fig:Diattenuation_Sim_vs_Exp}c are highlighted in the respective colours: bundle of densely grown fibres with \{$r \in [0.5,0.8]\,\um$, $t_{\text{sheath}} = 0.35\,r$, $n_{\text{m}} = 1.47$\} (orange triangle), bundle with broad distribution of radii $r \in [0.3,1.0]\,\um$ (brown triangle), bundle with thick myelin sheaths $t_{\text{sheath}} = 0.6\,r$ (green star), bundle with large fibres $r \in [2.5,4.0]\,\um$ (magenta star), bundle with reduced myelin refractive index $n_{\text{m}} = 1.39$ (blue circle). The mean diattenuation $\overline{\DS}$ becomes less negative with smaller fibre radius distribution, smaller myelin sheath thickness, larger fibre size (scaling factor), and smaller myelin refractive index.}
	\label{fig:Sim_Diattenuation_Parameters}
\end{figure}

\begin{multicols}{2} \small{ 
		
The diattenuation is mostly negative (magenta) for flat fibres ($\alpha \leq 50^{\circ}$) and becomes more positive (green) for steep fibres ($\alpha > 60^{\circ}$).
The histograms in Supplementary \cref{fig:Sim_DiattenuationMaps} show that the diattenuation values are almost symmetrically distributed around the mean value ($-17\,\% \pm 19\,\%$ for $\alpha = 0^{\circ}$ and $-22\,\% \pm 12\,\%$ for $\alpha = 50^{\circ}$). Thus, the mean value $\overline{\DS}$ is a good parameter to describe the diattenuation images.

\Cref{fig:Diattenuation_Sim_vs_Exp}c shows the simulated diattenuation curves (mean value of $\DS$ plotted against $\alpha$) for both fibre bundles: The mean diattenuation $\overline{\DS}$ is negative for non-steep fibres ($\alpha \leq 60^{\circ}$), positive for steep fibres ($70^{\circ} < \alpha < 90^{\circ}$), and almost zero for vertical fibres. For the bundle with broad fibre orientation distribution (red curve), the values range from $\overline{\DS} \approx -12.4\,\%$ to $1.2\,\%$. For the bundle of densely grown fibres (orange curve), the range is much larger ($\overline{\DS} \approx -22\,\%$ to $10\,\%$). 

The fibres were modelled with radii $r \in [0.5, 0.8]\,\um$, consisting of an inner axon and a surrounding myelin sheath with thickness $\dmyelin = 0.35\,r$ (see \hyperref[methods]{Methods}).
To study how the diattenuation depends on the fibre properties, the bundle of densely grown fibres was also simulated for a broader distribution of fibre radii ($r \in [0.3, 1.0]\,\um$, brown curve), and the horizontal bundle ($\alpha = 0^{\circ}$) for thicker myelin sheaths ($\dmyelin = 0.6\,r$, green star), and larger fibres ($r \in [2.5, 4.0]\,\um$, magenta star). While different fibre radii yield similar diattenuation curves, thicker myelin sheaths lead to more negative and larger fibres to less negative diattenuation values for $\alpha = 0^{\circ}$ (see also Supplementary \cref{fig:Sim_Diattenuation_Parameters}). Hence, regions with strongly negative diattenuation
belong most likely to strongly myelinated, relatively small, straight and horizontal fibres.

To estimate the combined diattenuation effect of anisotropic scattering and absorption, \cref{fig:Diattenuation_Sim_vs_Exp}c shows the simulated diattenuation curves $\overline{\DS}$ in direct comparison to the dichroism $\DK > 0$ plotted against the fibre inclination angle $\alpha$ (see Supplementary \hyperref[note1]{Note1} for derivation):
For regions with non-steep fibres ($\alpha < 65^{\circ}$), both diattenuation of type $D^+$ and $D^-$ are observed, depending on whether $\DK > 0$ or $\DS < 0$ dominates. Regions with steep fibres ($65^{\circ} < \alpha < 90^{\circ}$) only show diattenuation of type $D^+$ because both $\DS$ and $\DK$ are positive. As expected, regions with vertical fibres ($\alpha = 90^{\circ}$) show small diattenuation values ($\vert\DS\vert, \vert\DK\vert \ll 1$).


\paragraph{Simulated dependence of diattenuation on embedding time.}
As mentioned above, scattering is expected to decrease with increasing embedding time, \ie the refractive indices of the different tissue components become more similar to each other. A possible explanation for this behaviour is that the surrounding glycerine solution (with refractive index of 1.37) soaks into the myelin sheaths and reduces their effective refractive index.
So far, the simulations were performed for a myelin refractive index of 1.47, corresponding to literature values of lipids/membranes \cite{vanManen2008}. To model the diattenuation of brain tissue with long embedding time, the bundle of densely grown fibres was simulated for a reduced myelin refractive index of 1.39 (see \cref{fig:Diattenuation_Sim_vs_Exp}c, blue curve): the strength of the simulated diattenuation signals is much less ($\overline{\DS} \approx -2.2\,\%$ to $1.6\,\%$). 
Supplementary \cref{fig:Sim_Diattenuation_Parameters}d shows that $\vert\DS\vert$ decreases with decreasing myelin refractive index.
This suggests that anisotropic scattering ($\DS$) decreases with increasing time after embedding the brain section. 
As dichroism ($\DK$) is expected to remain positive, the net observed diattenuation is expected to become more positive over time.


\section*{Discussion}

In previous work (\textsc{Menzel} \ea\ (2017) \cite{menzel2017}), we have shown that brain tissue exhibits two different types of diattenuation: in some regions, the light is minimally attenuated when it is polarised parallel to the nerve fibres ($D^+$), in others, it is maximally attenuated ($D^-$). 
Here, we investigated this effect both in experimental studies and simulations and demonstrated that it depends on nerve fibre orientation, tissue composition, and embedding time.

Our experimental studies show that the same diattenuation effects can be observed in brain sections from different species (mouse, rat, monkey, human) and at different optical resolutions (from 43\,\um\,/\,px to 1.8\,\um\,/\,px): regions with out-of-plane fibres show almost exclusively diattenuation of type $D^+$, while regions with in-plane fibres are more likely to show diattenuation of type $D^-$ (see \cref{fig:DiattenuationMaps_Coronal-Sagittal}). With increasing time after embedding the brain sections in glycerine solution, the fraction of regions with diattenuation of type $D^+$ increases (see \cref{fig:Diattenuation_vs_Time_Vervet}).

Using a combination of analytical modelling and FDTD simulations, we could explain these experimental observations.
The diattenuation caused by anisotropic absorption (dichroism $\DK$) was described by an analytical model (see equation (19) in Supplementary \hyperref[note1]{Note1}): according to this model, the dichroism decreases with increasing out-of-plane inclination angle of the nerve fibres, only causes positive diattenuation (type $D^+$), and does not depend on the time after embedding the brain section.
The diattenuation caused by anisotropic scattering ($\DS$) was simulated for various fibre bundles with different inclination angles: regions with flat fibres (with respect to the section plane) show diattenuation of type $D^-$ while regions with steep fibres show diattenuation of type $D^+$. The strength of the simulated diattenuation signal depends on tissue properties like fibre orientation distribution, fibre size, and myelin sheath thickness, and decreases with increasing embedding time (see \cref{fig:Diattenuation_Sim_vs_Exp}c). 

To directly compare our simulation results to the experimental observations, we evaluated the diattenuation maps of the coronal vervet brain section in \cref{fig:Diattenuation_vs_Time_Vervet}a in regions with different fibre inclinations (see coloured circles in \cref{fig:Diattenuation_Sim_vs_Exp}d and e). 
In order to better compare the type of diattenuation to the fibre inclination, the brain sections on the right separately show the $D^+$ and $D^-$ regions in the white matter. Regions with diattenuation of type $D^+$ and regions with fibre inclinations $> 65^{\circ}$ are shown in green, while regions with diattenuation of type $D^-$ are shown in magenta. 
In freshly embedded brain sections, regions with steep fibres ($\alphaP > 65^{\circ}$) show almost exclusively type $D^+$: nearly all green regions in \cref{fig:Diattenuation_Sim_vs_Exp}e are also green in \cref{fig:Diattenuation_Sim_vs_Exp}d (see red circles). 
Regions with lower fibre inclinations show both types $D^+$ and $D^-$ (see yellow circles).
Regions with flat fibre inclinations are most likely to show type $D^-$ (see cyan circles).

All these observations can be explained by the combined model of analytically computed and simulated diattenuation curves (the dashed vertical lines in \cref{fig:Diattenuation_Sim_vs_Exp}c mark the inclination angles of the evaluated regions).
It should be noted that the analytical model of dichroism only allows a qualitative description as it does not consider any details in the fibre configurations. In future work, a more complex model could be used to study the exact dependence of the diattenuation on the underlying tissue properties.

The dependence of the diattenuation signal on the time after embedding the brain sections could successfully be modelled by reducing the refractive index of the myelin sheaths: the simulations have shown that a reduced myelin refractive index leads to a reduced anisotropic scattering (cf.\ orange and blue curves in \cref{fig:Diattenuation_Sim_vs_Exp}c), \ie brain tissue with long embedding time shows a smaller fraction of regions with diattenuation of type $D^-$ (cf.\ \cref{fig:Diattenuation_Sim_vs_Exp}d). The same model was used in \textsc{Menzel} \ea\ (2018) \cite{menzel2018-1} to explain the increasing transparency of brain tissue samples with increasing embedding time (see also Supplementary \cref{fig:Diattenuation_vs_Time_Rat}a), demonstrating the validity of our model. A possible explanation for the equalisation of refractive indices is that the embedding glycerine solution soaks into the myelin sheaths of the nerve fibres and reduces the effective refractive index of myelin.

While the transmitted light intensity and diattenuation caused by anisotropic scattering are dominated by light scattering and decrease with increasing embedding time, birefringence and dichroism are mostly independent of the embedding time, \ie they are probably caused by molecular effects.
The finding that the dichroism is positive ($\DK > 0$) means that the absorption becomes maximal when the light is polarised in the y-direction ($I_{\text{y}} < I_{\text{x}}$, see equation (\ref{eq:DS})), \ie perpendicularly to the nerve fibre axis and in the plane of the lipid molecules in the myelin sheath. This suggests that the dichroism of brain tissue is mainly caused by the myelin lipids.

Our simulations show that the strength and type of diattenuation ($D^+$ or $D^-$) depend not only on the out-of-plane inclination angle of the nerve fibres in the investigated brain sections and on the time after tissue embedding, but also on other tissue properties like homogeneity, fibre size, or myelin sheath thickness (cf.\ \cref{fig:Diattenuation_Sim_vs_Exp}c and Supplementary \cref{fig:Sim_Diattenuation_Parameters}b,c). 
A region with many small fibres, for example, is expected to show a stronger negative diattenuation than a region with few large fibres (see Supplementary \cref{fig:Sim_Diattenuation_Parameters}c). This allows to distinguish regions with similar myelin densities, \ie similar birefringence (3D-PLI) signals, but different tissue composition. How the measured type of diattenuation is exactly related to the underlying tissue properties needs to be investigated in future studies.

In conclusion, we could show that the diattenuation of brain tissue provides image contrasts between different tissue types and can therefore be used in addition to other imaging modalities to learn more about the brain's nerve fibre architecture and tissue composition, and to identify (pathological) changes.
This makes Diattenuation Imaging a valuable imaging technique.

} 


{\footnotesize

\section*{Methods and Materials}
\label{methods}

\paragraph{\footnotesize Preparation of brain sections.}
The measurements were performed on healthy brains from different species: mouse (\textit{C57BL/6}, male, six months old), rat (\textit{Wistar}, male, three months old), vervet monkey (African green monkey: \textit{Chlorocebus aethiops sabaeus}, male, between one and two years old), and human (male, 87 years old).
All animal procedures were approved by the institutional animal welfare committee at Forschungszentrum Jülich GmbH, Germany, and were in accordance with European Union (National Institutes of Health) guidelines for the use and care of laboratory animals.
The human brain was acquired in accordance with the local ethic committee of the University of Rostock, Germany. A written informed consent of the subject is available.

The brains were removed from the skull within 24 hours after death, fixed with 4\,\% buffered formaldehyde for several weeks, immersed in solutions of 10\,\% and 20\,\% glycerine combined with 2\,\% Dimethyl sulfoxide for cryo-protection, and deeply frozen. The frozen brains were cut with a cryostat microtome (\textit{Leica Microsystems}, Germany) into sections of 60\,\textmu m. This section thickness ensures both a good quality of the sections (\ie reduced fissures/deformations) and a sufficient signal strength \cite{MAxer2011_1,MAxer2011_2}. The brain sections were mounted on glass slides, embedded in a solution of 20\,\% glycerine, and cover-slipped. The glycerine solution can be used both as cryoprotectant and embedding medium because it does not impair the polarimetric measurements. The brain sections were measured one day after embedding (freshly embedded sections) or after several days to study the dependence of diattenuation on the embedding time. 


\paragraph{\footnotesize Polarimetric measurements.}
The measurements were performed with the in-house developed \textit{Large-Area Polarimeter (LAP)} which was also used for the diattenuation studies by \textsc{Menzel} \ea\ (2017) \cite{menzel2017}. 
The LAP consists of a green light source, a pair of crossed linear polarisers, and a quarter-wave retarder and specimen stage mounted in between the polarisers (see \cref{fig:DI}a).
The light source contains a matrix of $36 \times 36$ LEDs (\textit{NSPG 510S}, \textit{Nichia corporation}) and a diffuser plate (PMMA) and emits mostly incoherent and unpolarised light with a wavelength $\lambda = (525 \pm 25)$\,nm. The polarisers (\textit{XP38}) and the retarder (\textit{WP140HE}) were manufactured by \textit{ITOS}, Germany. During a measurement, the polarisers and the retarder were rotated simultaneously by angles $\rho = \{0^{\circ}, 10^{\circ}, \dots, 170^{\circ}$\} and the transmitted light intensity $I(\rho)$ was recorded by a CCD camera (\textit{AxioCam HRc} by \textit{Zeiss}) with the microscanning procedure of the camera sensor, yielding 4164 $\times$ 3120 pixels with a resolution down to 14\,\um\,/\,px.

The in-plane orientation angles $\phiP$ of the nerve fibres were obtained from 3D-PLI measurements \cite{MAxer2011_1,MAxer2011_2} with the above setup and by performing a discrete harmonic Fourier analysis on the measured light intensities per image pixel ($I_{\text{P}}(\rho) = a_{0\text{P}} + a_{2\text{P}}\,\cos(2\rho) + b_{2\text{P}}\,\sin(2\rho)$) and evaluating the phase of the signal: $\phiP = \atantwo(-a_{2\text{P}}, b_{2\text{P}})/2$. 
The amplitude of the signal $|\sin\deltaP|$ is related to the phase shift $\deltaP$ induced by the birefringent brain section \cite{menzel2015} ($|\sin\deltaP| = (a_{2\text{P}}^2 + b_{2\text{P}}^2)^{1/2} / a_{0\text{P}}, \,\,\, \deltaP \propto \cos^2\alphaP$) and was used to compute the out-of-plane inclination angles $\alphaP$ of the fibres in \cref{fig:Diattenuation_Sim_vs_Exp}e, making use of a tiltable specimen stage \cite{wiese2014} to improve the determination of the inclination angles.

The diattenuation measurements \cite{menzel2017} were performed by removing the retarder and the second polariser from the light path and measuring the transmitted light intensities as described above (see \cref{fig:DI}d). To account for the lower signal-to-noise ratio, the image of the brain section was recorded 20 times for each filter position and averaged. The strength of the measured diattenuation signal $\Dm$ and the phase $\phiD$ were determined from the amplitude and phase of the resulting light intensity profile ($I_{\text{D}}(\rho)  = a_{0\text{D}} + a_{2\text{D}}\,\cos(2\rho) + b_{2\text{D}}\,\sin(2\rho)$): $\Dm = (a_{2\text{D}}^2 + b_{2\text{D}}^2)^{1/2}/a_{0\text{D}}$ and $\phiD = \atantwo(b_{2\text{D}}, a_{2\text{D}})/2$.

The Fourier coefficients of order zero $\{a_{0\text{P}}, a_{0\text{D}}\}$ were used to register the images of the 3D-PLI measurements onto the images of the diattenuation measurements using in-house developed software tools based on the software packages \textit{ITK}, \textit{elastix}, and \textit{ANTs} \cite{elastix,shamonin2013,avants2008,avants2011,itk} which perform linear and non-linear transformations.

The resulting images $\{\phiP, \phiD, \Dm\}$ were used to generate the diattenuation images (see \cref{fig:DI}c): $\Dm$ values belonging to regions with $(\phiD - \phiP) \in [-20^{\circ},20^{\circ}]$ were colourised in green (referred to as $D^+$), regions with $(\phiD - \phiP) \in [-70^{\circ},110^{\circ}]$ were colourised in magenta (referred to as $D^-$). The angle ranges account for the uncertainties of $\phiD$ due to the non-ideal optical properties of the LAP \cite{menzel2017}: max$|\phiD-\phiP| \approx 20^{\circ} (\pm 90^{\circ})$ for $\Dm = 1\,\%$.


\paragraph{\footnotesize Measurements with prototypic polarising microscope.}
The prototypic polarising microscope has been described by \textsc{Wiese} \ea\ (2016) \cite{wiese}. It consists of a green light source ($\lambda = (532 \pm 5)$\,nm), a rotatable polariser, a fixed circular analyser (quarter-wave retarder with linear polariser), and a CCD camera. The polarising filters are of higher quality than in the LAP. The microscope objective has a $4 \times$ magnification and a numerical aperture of 0.2, yielding a pixel size in object space of about 1.8\,\um.
The measurements with the prototypic polarising microscope (see Supplementary \cref{fig:DiattenuationMaps_Taorad}) were performed as described above. Due to a superior signal-to-noise ratio, repeated diattenuation measurements were not necessary.


\paragraph{\footnotesize Model of nerve fibre configurations.}
The fibre bundles used for the FDTD simulations (cf.\ \cref{fig:Diattenuation_Sim_vs_Exp}a) were generated by in-house developed software similar to software used in other groups \cite{altendorf2010} allowing to create densely packed fibres without intersections: 700 straight fibres with different radii were randomly uniformly placed in an area of $45 \times 30$\,\um$^2$ and divided into segments of 2--5\,\um. The fibre segments were assigned a random displacement in x,y,z: max.\ 10\,\um\ for the bundle with broad fibre orientation distribution (\cref{fig:Diattenuation_Sim_vs_Exp}a, left) and max.\ 1\,\um\ for the bundle of densely grown fibres (\cref{fig:Diattenuation_Sim_vs_Exp}a, right). The resulting fibre segments were split or merged until the length of each segment was again between 2--5\,\um, ensuring that the maximum angle between adjacent segments was less than $20^{\circ}$. When a collision between two segments was detected, the segments were exposed to a small repelling force and the previous step was repeated until no more collisions were detected.
The mode angle difference between the local fibre orientation vectors and the predominant orientation of the resulting fibre bundle is about $25^{\circ}$ for the bundle with broad fibre orientation distribution and less than $10^{\circ}$ for the bundle of densely grown fibres.
To generate fibre bundles with different inclination angles, the bundles were rotated around the y-axis with respect to the center position and cropped to a volume of $30 \times 30 \times 30 \,\um^3$. To prevent fibres from touching each other after discretisation, all diameters were reduced by 5\,\%.

Each fibre with radius $r$ was modelled by an inner axon ($r_{\text{ax}} = 0.65\,r$) and a surrounding myelin sheath with thickness $t_{\text{sheath}} = 0.35\,r$, which contributes approximately one third to the overall fibre radius \cite{morell}. 
In brain tissue, the myelin sheath consists of densely packed cell membranes, \ie alternating layers of lipid bilayers (about 5\,nm thickness) and intra- or extracellular space (about 3\,nm thickness) \cite{martenson,leeH2014}. When the brain sections are embedded in glycerine solution (as described above), the extracellular space is expected to become filled with glycerine solution and increases \cite{quarles2006,martenson}.
The myelin (lipid) and glycerine layers were therefore assumed to contribute 3/4 and 1/4 to the overall myelin sheath thickness, respectively.
The simulations were performed for a simplified nerve fibre model, consisting of two myelin layers ($t_{\text{m}} = 3/7\,t_{\text{sheath}}$) and a separating glycerine layer ($t_{\text{g}} = 1/7\,t_{\text{sheath}}$).
The refractive indices of axon, myelin and glycerine layers were estimated from literature values of cytoplasm\cite{duck1990} ($n_{\text{ax}} = 1.35$), lipids/membranes\cite{vanManen2008} ($n_{\text{m}} = 1.47$), and from refractive index measurements of the glycerine solution used for embedding the brain sections ($n_{\text{g}} = 1.37$, measured with digital refractometer).
The surrounding medium was assumed to be homogeneous with a refractive index $n_{\text{surr}} = 1.37$, corresponding to the refractive index of gray matter \cite{sun2012}.
The absorption coefficients of brain tissue are small \cite{schwarzmaier1997,yaroslawsky2002} and were therefore neglected.
More details about the nerve fibre model can be found in \textsc{Menzel} \ea\ (2018) \cite{menzel2018-1}.

The fibres shown in \cref{fig:Diattenuation_Sim_vs_Exp}a have radii between $0.5\,\um$ and $0.8\,\um$ and a myelin sheath thickness of $0.35\,r$. 
The bundle of densely grown fibres was also simulated for different values of $r$, $\dmyelin$, and $\nmyelin$ (see Figs.\ \ref{fig:Diattenuation_Sim_vs_Exp}c and Supplementary \cref{fig:Sim_Diattenuation_Parameters}).


\paragraph{\footnotesize FDTD simulations.}
The propagation of the polarised light wave through the sample was computed by \textit{TDME3D}$^{\text{TM}}$ -- a massively parallel three-dimensional Maxwell Solver \cite{michielsen2010, wilts2014,wilts2012} based on a finite-difference time-domain (FDTD) algorithm \cite{taflove,deRaedt}. The algorithm numerically computes the electromagnetic field components by discretising space and time and approximating Maxwell's curl equations by finite differences \cite{menzel2016}.

The sample contains the fibre configuration ($30 \times 30 \times 30\,\um^3$, see above) and $0.5\,\um$ thick layers of glycerine solution on top and at the bottom.
The dimensions of the simulation volume are $30 \times 30 \times 35\,\um^3$, including 1\,\um\ thick uniaxial perfectly matched layer absorbing boundaries. The different components of the sample were modelled as dielectrics with real refractive indices $n$ (defined above).
The sample was illuminated by a plane monochromatic wave with linear polarisation (along the x- or y-axis).
The simulations were performed on the supercomputer \textit{JUQUEEN} \cite{juqueen} at Forschungszentrum Jülich GmbH, Germany, with 200 periods, a Courant factor of $0.8$, and a Yee mesh size of 25\,nm.
Using an MPI grid of $16 \times 16 \times 16$, each simulation run (calculation of one configuration and one polarisation state) consumed between 7000--8000 core hours, required a minimum memory between 260--360\,GB, and lasted between 1:45--2:00 hours.

The polarimeter used for the polarimetric measurements (LAP) has a broad wavelength spectrum and emits mostly incoherent light under different angles of incidence. As each simulation run is performed with coherent light with a certain wavelength and angle of incidence, the LAP is not well suited to be modelled by FDTD simulations.
Instead, we performed the simulations for the imaging system of the \textit{Polarising Microscope (PM)} \cite{MAxer2011_1,MAxer2011_2}, which has also been used to model the 3D-PLI measurement \cite{menzel2018-1}. 
The PM (manufactured by \textit{Taorad GmbH}, Germany) contains higher quality components and uses more coherent and less diffusive light with $\lambda = (550 \pm 5)\,$nm. As the polarising filters cannot be removed, the PM cannot be used for diattenuation measurements. However, the optics of the PM are similar to those of the prototypic polarising microscope (the microscope objective has a $5 \times$ magnification, a numerical aperture of 0.15, and yields an object-space resolution of 1.33\,\um\,/\,px). Therefore, the diattenuation effects simulated for the optics of the PM are presumably similar to those observed in the measurements (cf.\ Supplementary \cref{fig:DiattenuationMaps_Taorad}).

The imaging system of the PM was simulated as described by \textsc{Menzel} \ea\ (2018) \cite{menzel2018-1}:
The simulations were performed for normally incident light with 550\,nm wavelength. The propagation of the light wave through the sample (fibre configuration) was computed by TDME3D, yielding a superposition of monochromatic plane waves with different wave vectors.
The numerical aperture (NA $\approx 0.15$) was modelled by considering only wave vector angles $\leq \arcsin(\text{NA}) \approx 8.6^{\circ}$.
The spherical microlenses of the camera sensor were modelled by applying a moving average over the area of the microlens with a diameter of 1.33\,\um.
The light intensity recorded by the camera was computed as the absolute squared value of the electric field vector (neglecting angle dependencies because of the small aperture). To obtain the intensity at a certain point in the image plane, the electric field vectors for different wave vectors were summed and averaged over time. 
More details about the simulation model and error analysis can be found in \textsc{Menzel} \ea\ (2018) \cite{menzel2018-1}.

To model the diattenuation measurement, the simulations were performed for light polarised along the x-axis and along the y-axis. The resulting light intensities ($I_{\text{x}}$ and $I_{\text{y}}$) were used to compute the diattenuation images: $\DS = (I_{\text{x}} - I_{\text{y}})/(I_{\text{x}} + I_{\text{y}})$. 
Positive diattenuation values ($\DS > 0 \Leftrightarrow I_{\text{x}} > I_{\text{y}}$) correspond to regions with $D^+$ effect and were shown in green, regions with negative values ($\DS < 0 \Leftrightarrow I_{\text{x}} < I_{\text{y}}$) correspond to regions with $D^-$ effect and were shown in magenta (cf.\ \cref{fig:Diattenuation_Sim_vs_Exp}b and Supplementary \cref{fig:Sim_DiattenuationMaps}). 


\paragraph{\footnotesize Code and data availability.}
Image processing and data analysis were performed with \textit{Fiji} (https://fiji.sc/Fiji) and in-house developed software tools. The artificial fibre geometries were generated with in-house developed software (using \textit{Python} 3 and \textit{C++}) similar to software developed by Altendorf \& Jeulin \cite{altendorf2010}.
The electric field vectors obtained from the Maxwell Solver were processed with in-house developed software using \textit{Python} 2.7 and \textit{Numpy} 1.12.
All relevant program code, including the geometry parameters of the modelled fibre configurations, are available from the corresponding author upon reasonable request for purposes of academic research.
For the FDTD simulations, we used TDME3D, a massively parallel Maxwell solver \cite{michielsen2010, wilts2014,wilts2012}. 
The software is property of EMBD (European Marketing and Business Development BVBA).
All data supporting the findings of this study are available from the corresponding author on reasonable request. 


\bibliographystyle{unsrtnat}
\bibliography{BIBLIOGRAPHY}

\begin{thebibliography}{62}
\providecommand{\natexlab}[1]{#1}
\providecommand{\url}[1]{\texttt{#1}}
\expandafter\ifx\csname urlstyle\endcsname\relax
  \providecommand{\doi}[1]{doi: #1}\else
  \providecommand{\doi}{doi: \begingroup \urlstyle{rm}\Url}\fi

\bibitem[Lubetzki and Stankoff(2014)]{lubetzki2014}
C.~Lubetzki and B.~Stankoff.
\newblock Chapter 4 - demyelination in multiple sclerosis.
\newblock In D.~S. Goodin, editor, \emph{Multiple Sclerosis and Related
  Disorders}, volume 122 of \emph{Handbook of Clinical Neurology}, pages
  89--99. Elsevier, 2014.
\newblock \doi{10.1016/B978-0-444-52001-2.00004-2}.

\bibitem[Wenning et~al.(2008)Wenning, Stefanova, Jellinger, Poewe, and
  Schlossmacher]{wenning2008}
Gregor~K. Wenning, Nadia Stefanova, Kurt~A. Jellinger, Werner Poewe, and
  Michael~G. Schlossmacher.
\newblock Multiple system atrophy: {A} primary oligodendrogliopathy.
\newblock \emph{Annals of Neurology}, 64\penalty0 (3):\penalty0 239--246, 2008.
\newblock \doi{10.1002/ana.21465}.

\bibitem[Minnerop et~al.(2010)Minnerop, Lüders, Specht, Ruhlmann, Schimke,
  Thompson, Chou, Toga, Abele, W\"{u}llner, and Klockgether]{minnerop2010}
M.~Minnerop, E.~Lüders, K.~Specht, J.~Ruhlmann, N.~Schimke, P.~M. Thompson,
  Y.~Y. Chou, A.~W. Toga, M.~Abele, U.~W\"{u}llner, and T.~Klockgether.
\newblock Callosal tissue loss in multiple system atrophy--a one-year follow-up
  study.
\newblock \emph{Movement disorders}, 25\penalty0 (15):\penalty0 2613--2620,
  2010.

\bibitem[Ferrer(2018)]{ferrer2018}
I.~Ferrer.
\newblock Oligodendrogliopathy in neurodegenerative diseases with abnormal
  protein aggregates: {T}he forgotten partner.
\newblock \emph{Progress in Neurobiology}, 169:\penalty0 24--54, 2018.
\newblock ISSN 0301-0082.
\newblock \doi{10.1016/j.pneurobio.2018.07.004}.

\bibitem[van~der Knaap and Bugiani(2017)]{knaap2017}
M.~S. van~der Knaap and M.~Bugiani.
\newblock Leukodystrophies: a proposed classification system based on
  pathological changes and pathogenetic mechanisms.
\newblock \emph{Acta neuropathologica}, 134\penalty0 (3):\penalty0 351--382,
  2017.
\newblock \doi{10.1007/s00401-017-1739-1}.

\bibitem[Glasser and Van~Essen(2011)]{glasser2011}
Matthew~F. Glasser and David~C. Van~Essen.
\newblock Mapping human cortical areas in vivo based on myelin content as
  revealed by {T1}- and {T2}-weighted {MRI}.
\newblock \emph{Journal of Neuroscience}, 31\penalty0 (32):\penalty0
  11597--11616, 2011.
\newblock ISSN 0270-6474.
\newblock \doi{10.1523/JNEUROSCI.2180-11.2011}.

\bibitem[Mori and Zhang(2006)]{mori2006}
S.~Mori and J.~Zhang.
\newblock Principles of diffusion tensor imaging and its applications to basic
  neuroscience research.
\newblock \emph{Neuron}, 51\penalty0 (5):\penalty0 527--539, 2006.
\newblock \doi{https://doi.org/10.1016/j.neuron.2006.08.012}.

\bibitem[Tuch et~al.(2003)Tuch, Reese, Wiegell, and Wedeen]{tuch2003}
D.~S. Tuch, T.~G. Reese, M.~R. Wiegell, and V.~J. Wedeen.
\newblock Diffusion {MRI} of complex neural architecture.
\newblock \emph{Neuron}, 40\penalty0 (5):\penalty0 885--895, 2003.
\newblock \doi{https://doi.org/10.1016/S0896-6273(03)00758-X}.

\bibitem[Chang et~al.(2015)Chang, Sundman, Petit, Guhaniyogi, Chu, Petty, Song,
  and Chen]{chang2015}
H.-C. Chang, M.~Sundman, L.~Petit, S.~Guhaniyogi, M.-L. Chu, C.~Petty, A.~W.
  Song, and N.~Chen.
\newblock Human brain diffusion tensor imaging at submillimeter isotropic
  resolution on a 3 {T}esla clinical {MRI} scanner.
\newblock \emph{NeuroImage}, 118:\penalty0 667--675, 2015.
\newblock \doi{https://doi.org/10.1016/j.neuroimage.2015.06.016}.

\bibitem[Zeineh et~al.(2016)Zeineh, Palomero-Gallagher, Axer, Gr\"{a}{\ss}el,
  Goubran, Wree, Woods, Amunts, and Zilles]{zeineh2016}
M.~Zeineh, N.~Palomero-Gallagher, M.~Axer, D.~Gr\"{a}{\ss}el, M.~Goubran,
  A.~Wree, R.~Woods, K.~Amunts, and K~Zilles.
\newblock Direct visualization and mapping of the spatial course of fiber
  tracts at microscopic resolution in the human hippocampus.
\newblock \emph{Cerebral Cortex}, 2016.
\newblock \doi{https://doi.org/10.1093/cercor/bhw010}.

\bibitem[Henssen et~al.(2018)Henssen, Mollink, Kurt, van Dongen, Bartels,
  Gr{\"a}$\beta$el, Kozicz, Axer, and Van Cappellen~van Walsum]{henssen2018}
Dylan J. H.~A. Henssen, Jeroen Mollink, Erkan Kurt, Robert van Dongen, Ronald
  H. M.~A. Bartels, David Gr{\"a}$\beta$el, Tamas Kozicz, Markus Axer, and
  Anne-Marie Van Cappellen~van Walsum.
\newblock Ex vivo visualization of the trigeminal pathways in the human
  brainstem using 11.7{T} diffusion {MRI} combined with microscopy polarized
  light imaging.
\newblock \emph{Brain Structure and Function}, Oct 2018.
\newblock ISSN 1863-2661.
\newblock \doi{10.1007/s00429-018-1767-1}.
\newblock URL \url{https://doi.org/10.1007/s00429-018-1767-1}.

\bibitem[Caspers and Axer(2017)]{caspers2017}
S.~Caspers and M.~Axer.
\newblock Decoding the microstructural correlate of diffusion {MRI}.
\newblock \emph{NMR in Biomedicine}, page e3779, 2017.
\newblock \doi{10.1002/nbm.3779}.

\bibitem[Axer et~al.(2011{\natexlab{a}})Axer, Amunts, Gr{\"a}ssel, Palm,
  Dammers, Axer, Pietrzyk, and Zilles]{MAxer2011_1}
M.~Axer, K.~Amunts, D.~Gr{\"a}ssel, C.~Palm, J.~Dammers, H.~Axer, U.~Pietrzyk,
  and K.~Zilles.
\newblock A novel approach to the human connectome: Ultra-high resolution
  mapping of fiber tracts in the brain.
\newblock \emph{NeuroImage}, 54\penalty0 (2):\penalty0 1091--1101,
  2011{\natexlab{a}}.
\newblock \doi{https://doi.org/10.1016/j.neuroimage.2010.08.075}.

\bibitem[Axer et~al.(2011{\natexlab{b}})Axer, Gr{\"a}ssel, Kleiner, Dammers,
  Dickscheid, Reckfort, H{\"u}tz, Eiben, Pietrzyk, Zilles, and
  Amunts]{MAxer2011_2}
M.~Axer, D.~Gr{\"a}ssel, M.~Kleiner, J.~Dammers, T.~Dickscheid, J.~Reckfort,
  T.~H{\"u}tz, B.~Eiben, U.~Pietrzyk, K.~Zilles, and K.~Amunts.
\newblock High-resolution fiber tract reconstruction in the human brain by
  means of three-dimensional polarized light imaging.
\newblock \emph{Frontiers in Neuroinformatics}, 5\penalty0 (34):\penalty0
  1--13, 2011{\natexlab{b}}.
\newblock \doi{https://doi.org/10.3389/fninf.2011.00034}.

\bibitem[Schmitt and Bear(1939)]{schmitt1939}
F.~O. Schmitt and R.~S. Bear.
\newblock The ultrastructure of the nerve axon sheath.
\newblock \emph{Biological reviews of the Cambridge Philosophical Society},
  14:\penalty0 27--50, 1939.

\bibitem[Koike-Tani et~al.(2013)Koike-Tani, Tani, Mehta, Verma, and
  Oldenbourg]{koike-tani2013}
M.~Koike-Tani, T.~Tani, S.~B. Mehta, A.~Verma, and R.~Oldenbourg.
\newblock Polarized light microscopy in reproductive and developmental biology.
\newblock \emph{Molecular Reproduction and Development}, pages 1--15, 2013.
\newblock ISSN 1098-2795.
\newblock \doi{https://doi.org/10.1002/mrd.22221}.

\bibitem[Menzel et~al.(2015)Menzel, Michielsen, De~Raedt, Reckfort, Amunts, and
  Axer]{menzel2015}
M.~Menzel, K.~Michielsen, H.~De~Raedt, J.~Reckfort, K.~Amunts, and M.~Axer.
\newblock A {J}ones matrix formalism for simulating three-dimensional polarized
  light imaging of brain tissue.
\newblock \emph{Journal of the Royal Society Interface}, 12:\penalty0 20150734,
  2015.
\newblock \doi{https://doi.org/10.1098/rsif.2015.0734}.

\bibitem[Martenson(1992)]{martenson}
R.~E. Martenson.
\newblock \emph{Myelin: Biology and Chemistry}.
\newblock CRC Press, USA, 1992.
\newblock ISBN 0-8493-8849-X.

\bibitem[Mehta et~al.(2013)Mehta, Shribak, and Oldenbourg]{mehta2013}
S.~B. Mehta, M.~Shribak, and R.~Oldenbourg.
\newblock Polarized light imaging of birefringence and diattenuation at high
  resolution and high sensitivity.
\newblock \emph{Journal of Optics}, 15:\penalty0 1--13, 2013.
\newblock \doi{https://doi.org/10.1088/2040-8978/15/9/094007}.

\bibitem[Chenault and Chipman(1993)]{chenault1993}
D.~B. Chenault and R.~A. Chipman.
\newblock Measurements of linear diattenuation and linear retardance spectra
  with a rotating sample spectropolarimeter.
\newblock \emph{Applied Optics}, 32\penalty0 (19):\penalty0 3513--3519, 1993.
\newblock \doi{https://doi.org/10.1364/AO.32.003513}.

\bibitem[Ghosh and Vitkin(2011)]{ghosh2011}
N.~Ghosh and I.~A. Vitkin.
\newblock Tissue polarimetry: concepts, challenges, applications, and outlook.
\newblock \emph{Journal of Biomedical Optics}, 16\penalty0 (11):\penalty0
  110801, 2011.
\newblock \doi{https://doi.org/10.1117/1.3652896}.

\bibitem[Chipman(1994)]{chipman}
R.~A. Chipman.
\newblock Polarimetry.
\newblock In \emph{Handbook of Optics, Vol. 2: Devices, Measurements, and
  Properties}, chapter~22, pages 22.1--22.37. McGraw-Hill, New York, 2 edition,
  1994.
\newblock ISBN 978-0070479746.

\bibitem[Soni et~al.(2013)Soni, Purwar, Lakhotia, Chandel, Banerjee, Kumar, and
  Ghosh]{soni2013}
J.~Soni, H.~Purwar, H.~Lakhotia, S.~Chandel, C.~Banerjee, U.~Kumar, and
  N.~Ghosh.
\newblock Quantitative fluorescence and elastic scattering tissue polarimetry
  using an {E}igenvalue calibrated spectroscopic {M}ueller matrix system.
\newblock \emph{Optics Express}, 21\penalty0 (13):\penalty0 15475--15489, 2013.
\newblock \doi{https://doi.org/10.1364/OE.21.015475}.

\bibitem[Jiao et~al.(2003)Jiao, Yu, Stoica, and Wang]{jiao2003}
S.~Jiao, W.~Yu, G.~Stoica, and L.~V. Wang.
\newblock Contrast mechanisms in polarization-sensitive {M}ueller-matrix
  optical coherence tomography and application in burn imaging.
\newblock \emph{Applied Optics}, 42\penalty0 (25):\penalty0 5191--5197, 2003.
\newblock \doi{https://doi.org/10.1364/AO.42.005191}.

\bibitem[Fan and Yao(2013)]{fan2013}
C.~Fan and G.~Yao.
\newblock Imaging myocardial fiber orientation using polarization sensitive
  optical coherence tomography.
\newblock \emph{Biomedical Optics Express}, 4\penalty0 (3):\penalty0 460--465,
  2013.
\newblock \doi{https://doi.org/10.1364/BOE.4.000460}.

\bibitem[Park et~al.(2004)Park, Pierce, Cense, and de~Boer]{park2004}
B.~Hyle Park, M.~C. Pierce, B.~Cense, and J.~F. de~Boer.
\newblock Jones matrix analysis for a polarization-sensitive optical coherence
  tomography system using fiber-optic components.
\newblock \emph{Optics Letters}, 29\penalty0 (21):\penalty0 2512--2514, 2004.
\newblock \doi{https://doi.org/10.1364/OL.29.002512}.

\bibitem[Swami et~al.(2006)Swami, Manhas, Buddhiwant, Ghosh, Uppal, and
  Gupta]{swami2006}
M.~K. Swami, S.~Manhas, P.~Buddhiwant, N.~Ghosh, A.~Uppal, and P.~K. Gupta.
\newblock Polar decomposition of {3{\texttimes}3 Mueller} matrix: a tool for
  quantitative tissue polarimetry.
\newblock \emph{Optics Express}, 14\penalty0 (20):\penalty0 9324--9337, 2006.
\newblock \doi{https://doi.org/10.1364/OE.14.009324}.

\bibitem[Westphal et~al.(2016)Westphal, Kaltenbach, and Wicker]{westphal2016}
P.~Westphal, J.~M. Kaltenbach, and K.~Wicker.
\newblock Corneal birefringence measured by spectrally resolved {M}ueller
  matrix ellipsometry and implications for non-invasive glucose monitoring.
\newblock \emph{Biomedical Optics Express}, 7\penalty0 (4):\penalty0 1160--74,
  2016.
\newblock \doi{https://doi.org/10.1364/BOE.7.001160}.

\bibitem[Naoun et~al.(2005)Naoun, Dorr, All\'{e}, Sablon, and
  Benoit]{naoun2005}
O.~K. Naoun, V.~L. Dorr, P.~All\'{e}, J.-C. Sablon, and A.-M. Benoit.
\newblock Exploration of the retinal nerve fiber layer thickness by measurement
  of the linear dichroism.
\newblock \emph{Applied Optics}, 44\penalty0 (33):\penalty0 7074--7082, 2005.
\newblock \doi{https://doi.org/10.1364/AO.44.007074}.

\bibitem[Huang(2006)]{huang2006}
X.-R. Huang.
\newblock Polarization properties of the retinal nerve fiber layer.
\newblock \emph{Bulletin de la Soci\'{e}t\'{e} belge d'ophtalmologie},
  302:\penalty0 71--88, 2006.

\bibitem[Menzel et~al.(2017)Menzel, Reckfort, Weigand, K\"{o}se, Amunts, and
  Axer]{menzel2017}
M.~Menzel, J.~Reckfort, D.~Weigand, H.~K\"{o}se, K.~Amunts, and M.~Axer.
\newblock Diattenuation of brain tissue and its impact on {3D} polarized light
  imaging.
\newblock \emph{Biomedical Optics Express}, 8\penalty0 (7):\penalty0
  3163--3197, 2017.
\newblock \doi{https://doi.org/10.1364/BOE.8.003163}.

\bibitem[Paxinos and Watson(2007)]{paxinos2007}
G.~Paxinos and C.~Watson.
\newblock \emph{The Rat Brain in Stereotaxic Coordinates}.
\newblock Academic Press, 6 edition, 2007.

\bibitem[Woods et~al.(2011)Woods, Fears, Jorgensen, Fairbanks, Toga, and
  Freimer]{woods2011}
R.~P. Woods, S.~C. Fears, M.~J. Jorgensen, L.~A. Fairbanks, A.~W. Toga, and
  N.~B. Freimer.
\newblock A web-based brain atlas of the vervet monkey, chlorocebus aethiops.
\newblock \emph{NeuroImage}, 54\penalty0 (3):\penalty0 1872--1880, 2011.
\newblock \doi{https://doi.org/10.1016/j.neuroimage.2010.09.070}.

\bibitem[Zilles et~al.(2015)Zilles, Palomero-Gallagher, Gr\"{a}{\ss}el,
  Schl{\"o}mer, Cremer, Woods, Amunts, and Axer]{zilles2015}
K.~Zilles, N.~Palomero-Gallagher, D.~Gr\"{a}{\ss}el, P.~Schl{\"o}mer,
  M.~Cremer, R.~Woods, K.~Amunts, and M.~Axer.
\newblock High-resolution fiber and fiber tract imaging using polarized light
  microscopy in the human, monkey, rat, and mouse brain.
\newblock In Kathleen~S. Rockland, editor, \emph{Axons and Brain Architecture},
  chapter~18, pages 369--389. Elsevier Acadamic Press, San Diego, 2015.

\bibitem[Taflove and Hagness(2005)]{taflove}
A.~Taflove and S.~C. Hagness.
\newblock \emph{Computational Electrodynamics: The {F}inite-Difference
  {T}ime-Domain Method}.
\newblock Artech House, MA USA, 3 edition, 2005.
\newblock ISBN 1580538320.

\bibitem[{Menzel} et~al.(2016){Menzel}, {Axer}, {De Raedt}, and
  {Michielsen}]{menzel2016}
M.~{Menzel}, M.~{Axer}, H.~{De Raedt}, and K.~{Michielsen}.
\newblock {Finite-Difference Time-Domain Simulation for Three-Dimensional
  Polarized Light Imaging}.
\newblock In K.~Amunts, L.~Grandinetti, T.~Lippert, and N.~Petkov, editors,
  \emph{{Brain-Inspired Computing. BrainComp 2015. Lecture Notes in Computer
  Science}}, volume 10087, chapter~6, pages 73--85. Springer International
  Publishing, Cham, 2016.
\newblock \doi{https://doi.org/10.1007/978-3-319-50862-7\_6}.

\bibitem[Menzel et~al.(2018)Menzel, Axer, Raedt, Costantini, Silvestri, Pavone,
  Amunts, and Michielsen]{menzel2018-1}
M.~Menzel, M.~Axer, H.~De Raedt, I.~Costantini, L.~Silvestri, F.~S. Pavone,
  K.~Amunts, and K.~Michielsen.
\newblock Transmittance assisted interpretation of {3D} nerve fibre
  architectures.
\newblock \emph{{Preprint at https://arxiv.org/abs/1806.07157}}, 2018.
\newblock URL \url{https://arxiv.org/abs/1806.07157}.

\bibitem[van Manen et~al.(2008)van Manen, Verkuijlen, Wittendorp, Subramaniam,
  van~den Berg, Roos, and Otto]{vanManen2008}
H.-J. van Manen, P.~Verkuijlen, P.~Wittendorp, V.~Subramaniam, T.~K. van~den
  Berg, D.~Roos, and C.~Otto.
\newblock Refractive index sensing of green fluorescent proteins in living
  cells using fluorescence lifetime imaging microscopy.
\newblock \emph{Biophysical Letters}, 94\penalty0 (8):\penalty0 L67--69, 2008.
\newblock \doi{https://doi.org/10.1529/biophysj.107.127837}.

\bibitem[Wiese et~al.(2014)Wiese, Gr{\"a}ssel, Pietrzyk, Amunts, and
  Axer]{wiese2014}
H.~Wiese, D.~Gr{\"a}ssel, U.~Pietrzyk, K.~Amunts, and M.~Axer.
\newblock Polarized light imaging of the human brain: a new approach to the
  data analysis of tilted sections.
\newblock In D.~B. Chenault and D.~H. Goldstein, editors, \emph{SPIE
  Proceedings, Polarization: Measurement, Analysis, and Remote Sensing XI},
  volume 9099, 2014.

\bibitem[Klein et~al.(2010)Klein, Staring, Murphy, Viergever, and
  Pluim]{elastix}
S.~Klein, M.~Staring, K.~Murphy, M.~A. Viergever, and J.~P.~W. Pluim.
\newblock elastix: A toolbox for intensity-based medical image registration.
\newblock \emph{IEEE Transactions on Medical Imaging}, 29\penalty0
  (1):\penalty0 196--205, 2010.
\newblock \doi{https://doi.org/10.1109/TMI.2009.2035616}.

\bibitem[Shamonin et~al.(2013)Shamonin, Bron, Lelieveldt, Smits, Klein, and
  Staring]{shamonin2013}
D.~P. Shamonin, E.~E. Bron, B.~P.~F. Lelieveldt, M.~Smits, S.~Klein, and
  M.~Staring.
\newblock Fast parallel image registration on {CPU} and {GPU} for diagnostic
  classification of {A}lzheimer's disease.
\newblock \emph{Frontiers in Neuroinformatics}, 7:\penalty0 50, 2013.
\newblock \doi{https://doi.org/10.3389/fninf.2013.00050}.

\bibitem[Avants et~al.(2008)Avants, Epstein, Grossman, and Gee]{avants2008}
B.~B. Avants, C.~L. Epstein, M.~Grossman, and J.~C. Gee.
\newblock Symmetric diffeomorphic image registration with cross-correlation:
  Evaluating automated labeling of elderly and neurodegenerative brain.
\newblock \emph{Medical Image Analysis}, 12\penalty0 (1):\penalty0 26--41,
  2008.
\newblock \doi{https://doi.org/10.1016/j.media.2007.06.004}.

\bibitem[Avants et~al.(2011)Avants, Tustison, Song, Cook, Klein, and
  Gee]{avants2011}
B.~B. Avants, N.~J. Tustison, G.~Song, P.~A. Cook, A.~Klein, and J.~C. Gee.
\newblock A reproducible evaluation of {ANT}s similarity metric performance in
  brain image registration.
\newblock \emph{NeuroImage}, 54\penalty0 (3):\penalty0 2033--2044, 2011.
\newblock \doi{https://doi.org/10.1016/j.neuroimage.2010.09.025}.

\bibitem[itk()]{itk}
National library of medicine insight segmentation and registration toolkit
  ({ITK}), \url{https://itk.org/}.

\bibitem[Wiese(2016)]{wiese}
H.~Wiese.
\newblock \emph{Enhancing the Signal Interpretation and Microscopical Hardware
  Concept of {3D} {P}olarized {L}ight {I}maging}.
\newblock PhD thesis, University of Wuppertal, 2016.

\bibitem[Altendorf and Jeulin(2010)]{altendorf2010}
H.~Altendorf and D.~Jeulin.
\newblock Random walk based stochastic modeling of {3D} fiber systems.
\newblock \emph{Physical Review E : Statistical, Nonlinear, and Soft Matter
  Physics, American Physical Society}, 83\penalty0 (4), 2010.
\newblock \doi{https://doi.org/10.1103/PhysRevE.83.041804}.

\bibitem[Morell et~al.(1989)Morell, Quarles, and Norton]{morell}
P.~Morell, R.~H. Quarles, and W.~T. Norton.
\newblock Formation, structure, and biochemistry of myelin.
\newblock In G.~J. Siegel, editor, \emph{Basic Neurochemistry -- Molecular,
  Cellular, and Medical Aspects}, pages 109--136. Raven Press, New York, 4
  edition, 1989.
\newblock ISBN 0881673439.

\bibitem[Lee et~al.(2014)Lee, Park, Seo, Park, and Kim]{leeH2014}
H.~Lee, J.~H. Park, I.~Seo, S.-H. Park, and S.~Kim.
\newblock Improved application of the electrophoretic tissue clearing
  technology, {CLARITY}, to intact solid organs including brain, pancreas,
  liver, kidney, lung, and intestine.
\newblock \emph{BMC Developmental Biology}, 14:\penalty0 781, 2014.
\newblock \doi{https://doi.org/10.1186/s12861-014-0048-3}.

\bibitem[Quarles et~al.(2006)Quarles, Macklin, and Morell]{quarles2006}
R.~H. Quarles, W.~B. Macklin, and P.~Morell.
\newblock Myelin formation, structure and biochemistry.
\newblock In G.~Siegel, R.~W. Albers, S.~Brady, and D.~Price, editors,
  \emph{Basic Neurochemistry: Molecular, Cellular and Medical Aspects}, pages
  51--71. Elsevier Academic Press, Burlington, MA, 7 edition, 2006.
\newblock ISBN 0-12-088397-X.

\bibitem[Duck(1990)]{duck1990}
F.~A. Duck.
\newblock \emph{Physical Properties of Tissue: A Comprehensive Reference Book}.
\newblock Academic Press, San Diego, 1990.
\newblock ISBN 9780122228001.

\bibitem[Sun et~al.(2012)Sun, Lee, Wu, Sarntinoranont, and Xie]{sun2012}
J.~Sun, S.~J. Lee, L.~Wu, M.~Sarntinoranont, and H.~Xie.
\newblock Refractive index measurement of acute rat brain tissue slices using
  optical coherence tomography.
\newblock \emph{Optics Express}, 20\penalty0 (2):\penalty0 1084--1095, 2012.
\newblock \doi{https://doi.org/10.1364/OE.20.001084}.

\bibitem[Schwarzmaier et~al.(1997)Schwarzmaier, Yaroslavsky, Yaroslavsky,
  Goldbach, Kahn, Ulrich, Schulze, and Schober]{schwarzmaier1997}
H.-J. Schwarzmaier, A.~Yaroslavsky, I.~Yaroslavsky, T.~Goldbach, T.~Kahn,
  F.~Ulrich, P.~C. Schulze, and R.~Schober.
\newblock Optical properties of native and coagulated human brain structures.
\newblock \emph{SPIE}, 2970:\penalty0 492--499, 1997.
\newblock \doi{https://doi.org/10.1117/12.275082}.

\bibitem[Yaroslavsky et~al.(2002)Yaroslavsky, Schulze, Yaroslavsky, Schober,
  Ulrich, and Schwarzmaier]{yaroslawsky2002}
A.~N. Yaroslavsky, P.~C. Schulze, I.~V. Yaroslavsky, R.~Schober, F.~Ulrich, and
  H.-J. Schwarzmaier.
\newblock Optical properties of selected native and coagulated human brain
  tissues in vitro in the visible and near infrared spectral range.
\newblock \emph{Physics in Medicine and Biology}, 47:\penalty0 2059--2073,
  2002.
\newblock URL \url{http://stacks.iop.org/PMB/47/2059}.

\bibitem[Michielsen et~al.(2010)Michielsen, De~Raedt, and
  Stavenga]{michielsen2010}
K.~Michielsen, H.~De~Raedt, and D.~G. Stavenga.
\newblock Reflectivity of the gyroid biophotonic crystals in the ventral wing
  scales of the {G}reen {H}airstreak butterfly, \textit{{C}allophrys rubi}.
\newblock \emph{Journal of the Royal Society Interface}, 7:\penalty0 765--771,
  2010.
\newblock \doi{https://doi.org/10.1098/rsif.2009.0352}.

\bibitem[Wilts et~al.(2014)Wilts, Michielsen, De~Raedt, and
  Stavenga]{wilts2014}
B.~D. Wilts, K.~Michielsen, H.~De~Raedt, and D.~G. Stavenga.
\newblock Sparkling feather reflections of a bird-of-paradise explained by
  finite-difference time-domain modeling.
\newblock \emph{Proceedings of the National Academy of Sciences}, 2014.
\newblock \doi{https://doi.org/10.1073/pnas.1323611111}.

\bibitem[Wilts et~al.(2012)Wilts, Michielsen, Kuipers, De~Raedt, and
  Stavenga]{wilts2012}
B.~D. Wilts, K.~Michielsen, J.~Kuipers, H.~De~Raedt, and D.~G. Stavenga.
\newblock Brilliant camouflage: photonic crystals in the diamond weevil,
  {E}ntimus imperialis.
\newblock \emph{Proceedings of the Royal Society B: Biological Sciences},
  279\penalty0 (1738):\penalty0 2524--2530, 2012.
\newblock ISSN 0962-8452.
\newblock \doi{https://doi.org/10.1098/rspb.2011.2651}.

\bibitem[De~Raedt(2005)]{deRaedt}
H.~De~Raedt.
\newblock Advances in unconditionally stable techniques.
\newblock In A.~Taflove and S.~C. Hagness, editors, \emph{Computational
  Electrodynamics: The Finite-Difference Time-Domain Method}, chapter~18.
  Artech House, MA USA, 3 edition, 2005.
\newblock ISBN 1580538320.

\bibitem[{J\"{u}lich Supercomputing Centre}(2015)]{juqueen}
{J\"{u}lich Supercomputing Centre}.
\newblock {JUQUEEN}: {IBM} {B}lue {G}ene/{Q} {S}upercomputer {S}ystem at the
  {J}{\"u}lich {S}upercomputing {C}entre.
\newblock \emph{{J}ournal of large-scale research facilities}, 1,\,A1, 2015.
\newblock \doi{https://doi.org/10.17815/jlsrf-1-18}.

\bibitem[{J\"{u}lich Supercomputing Centre}(2016)]{jureca}
{J\"{u}lich Supercomputing Centre}.
\newblock {JURECA: General-purpose supercomputer at J\"{u}lich Supercomputing
  Centre}.
\newblock \emph{Journal of large-scale research facilities}, 2,\,A62, 2016.
\newblock \doi{https://doi.org/10.17815/jlsrf-2-121}.

\bibitem[Born and Wolf(2011)]{born}
M.~Born and E.~Wolf.
\newblock \emph{Principles of Optics -- {E}lectromagnetic Theory of
  Propagation, Interference and Diffraction of Light}.
\newblock Cambridge University Press, 7 edition, 2011.
\newblock ISBN 978-0-521-64222-4.

\bibitem[Beuthan et~al.(1996)Beuthan, Minet, Helfmann, Herrig, and
  M\"{u}ller]{beuthan1996}
J.~Beuthan, O.~Minet, J.~Helfmann, M.~Herrig, and G.~M\"{u}ller.
\newblock The spatial variation of the refractive index in biological cells.
\newblock \emph{Physics in Medicine and Biology}, 41\penalty0 (3):\penalty0
  369--382, 1996.

\bibitem[Demtr\"{o}der(2009)]{demtroeder}
W.~Demtr\"{o}der.
\newblock \emph{{E}xperimentalphysik 2 -- {E}lektrizit\"{a}t und {O}ptik}.
\newblock Springer-Verlag Berlin Heidelberg, 5 edition, 2009.
\newblock ISBN 978-3-540-68210-3.

\end{thebibliography}


\section*{Acknowledgements}
This project has received funding from the Helmholtz Association portfolio theme \textit{`Supercomputing and Modeling for the Human Brain'}, from the European Union's Horizon 2020 Research and Innovation Programme under Grant Agreement No.\ 720270 (HBP SGA1) and 785907 (HBP SGA2), and from the National Institutes of Health under grant agreements No.\ R01MH092311 and 5P40OD010965.
We gratefully acknowledge the computing time granted through JARA-HPC on the supercomputer \textit{JURECA} \cite{jureca} and \textit{JUQUEEN} \cite{juqueen} at Forschungszentrum Jülich (FZJ). We thank our colleagues from the Institute of Neuroscience and Medicine (INM-1), FZJ: Markus Cremer, Christian Rademacher, and Patrick Nysten for the preparation of the histological brain sections, Julia Reckfort, Hasan Köse, David Gräßel, Isabelle Mafoppa Fomat, and Philipp Schlömer for the polarimetric measurements, and Felix Matuschke for providing the algorithm to generate the fibre configurations. Furthermore, we thank Taorad GmbH for providing the optical bench and equipment for the high-resolution diattenuation measurements, Andreas Wree (Institute for Anatomy, University of Rostock) for the human brain sample, and Karl Zilles (INM-1) and Roger Woods (UCLA Brain Mapping Center, Los Angeles) for collaboration in the vervet brain project.

\section*{Author Contributions}
M.M.\ substantially contributed to the conception and design of the study as well as to the analysis and interpretation of the experimental and simulated data and to the theoretical considerations. She analysed the measurements, carried out the simulations, created the figures, and wrote the manuscript.
M.A.\ participated in the conception and design of the study, contributed to the analysis and interpretation of the data, and to the revision of the manuscript.
K.A.\ contributed to the anatomical content of the study and to the revision of the manuscript.
H.D.R.\ contributed to the interpretation of the simulated data, to the theoretical considerations, and to the revision of the manuscript.
K.M.\ participated in the design of the study, contributed to the interpretation of the simulated data, to the theoretical considerations, and to the revision of the manuscript. All authors read the final manuscript and gave approval for publication.


} 
\end{multicols}


\newpage
\section*{\large{Supplementary Information}}


\begin{figure}[H]
	\centering
	\includegraphics[width=\textwidth]{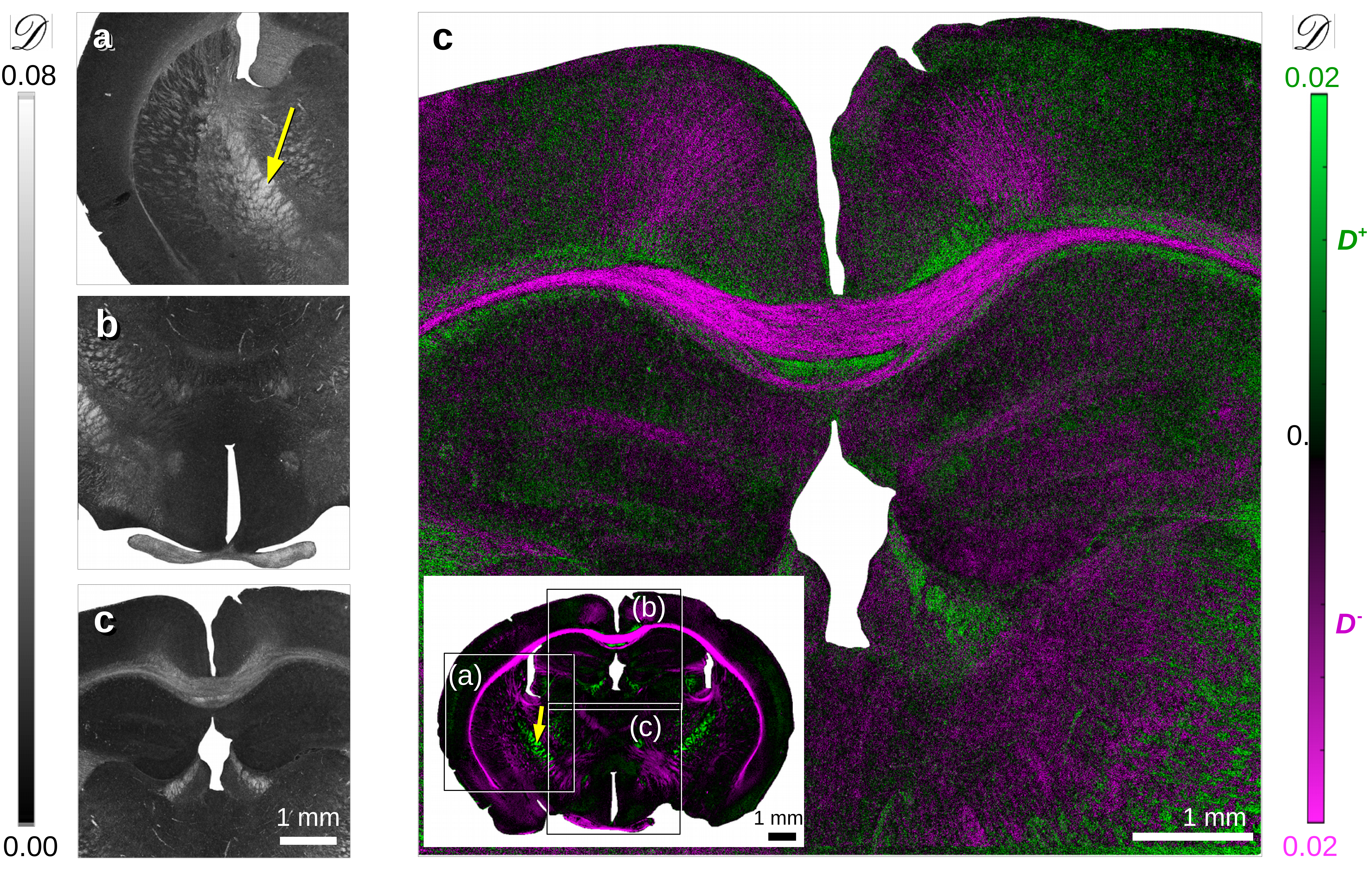}
	\caption{Diattenuation images of a coronal mouse brain section (cf.\ \cref{fig:DiattenuationMaps_Coronal-Sagittal}a) obtained from polarimetric measurements performed with the LAP (whole brain section, inset) and with a prototypic polarising microscope realised on an optical bench \cite{wiese} (enlarged areas a--c), see \hyperref[methods]{Methods}. The measurement with the LAP was performed with an effective object-space resolution of 14\,\um\,/\,px one day after tissue embedding. The measurements with the polarising microscope were performed with a pixel size of about 1.8\,\um\ two days after tissue embedding. The gray-scale images show the strength of the diattenuation signal $\Dm$, the coloured images show the diattenuation values belonging to regions with diattenuation of type $D^{+}$ ($\phiD - \phiP \in [-20^{\circ},20^{\circ}]$, in green) and regions with diattenuation of type $D^{-}$ ($\phiD - \phiP \in [70^{\circ},110^{\circ}]$, in magenta). The yellow arrows mark the maximum diattenuation measured with the microscope. The diattenuation in a region of $10 \times 10$ pixels ($\Dm \approx 9.9\,\%$) is much larger than the diattenuation of the same region measured with the LAP ($\Dm \approx 3.6\,\%$). However, large diattenuation values and diattenuation of type $D^+$ or $D^-$ are observed in similar regions, \ie the diattenuation effects do not depend on the optical resolution of the imaging system.}
	\label{fig:DiattenuationMaps_Taorad}
\end{figure}


\begin{figure}[H]
	\centering
	\includegraphics[width=0.95\textwidth]{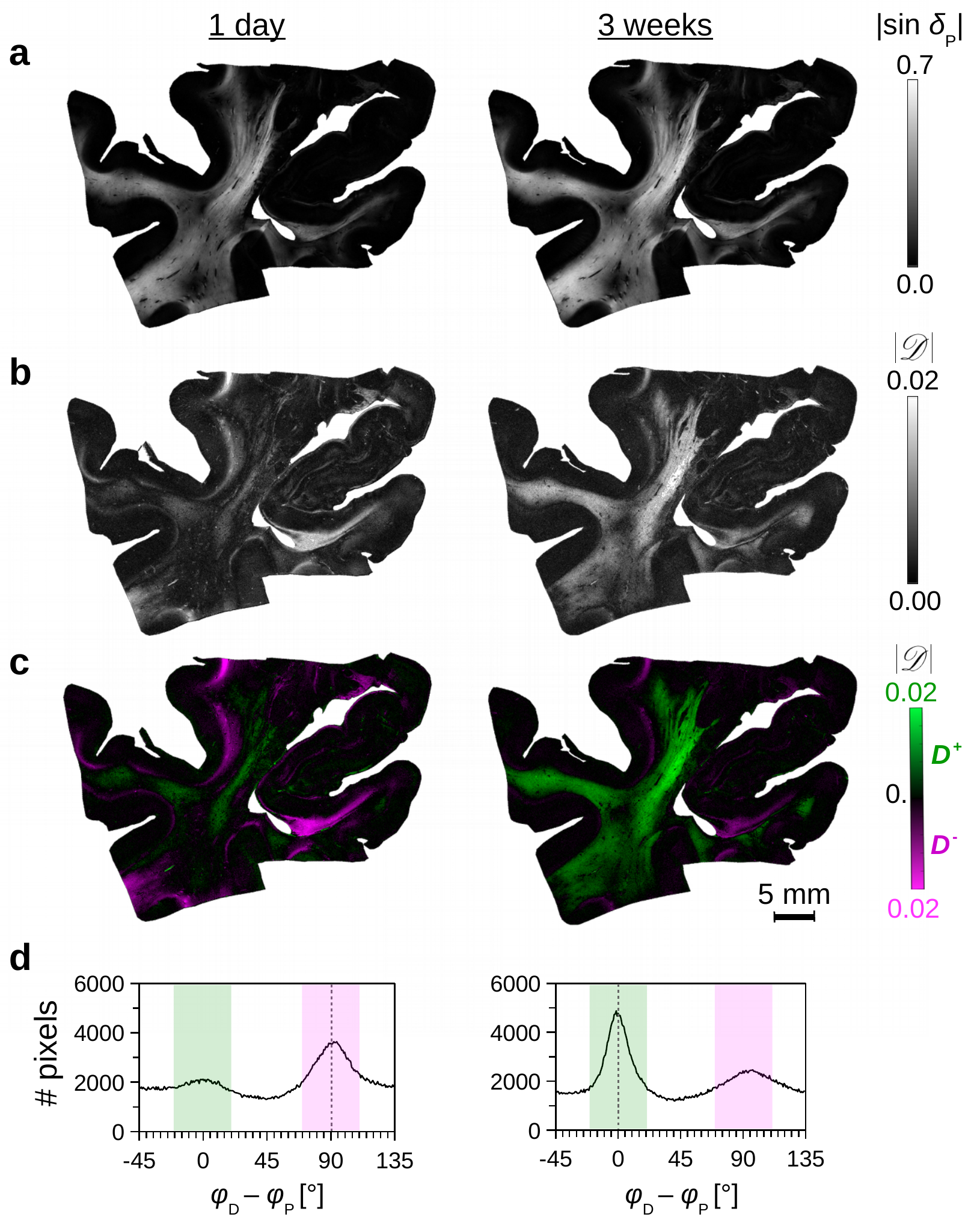}
	\caption{Dependence of diattenuation on embedding time for human brain tissue (coronal, 60 \um\ thick section of a human temporal lobe measured one day and three weeks after embedding the brain section). \textbf{a} Strength of the measured birefringence signal $\vert\sin\deltaP\vert$. \textbf{b} Strength of the measured diattenuation signal $\Dm$. \textbf{b--c} Diattenuation images and ($\phiD-\phiP$) histograms. Diattenuation values that belong to regions with diattenuation of type $D^{+}$ ($\phiD - \phiP \in [-20^{\circ},20^{\circ}]$) are shown in green, diattenuation values that belong to regions with diattenuation of type $D^{-}$ ($\phiD - \phiP \in [70^{\circ},110^{\circ}]$) are shown in magenta. The values $\{\Dm, \phiD, \phiP\}$ were determined from polarimetric measurements with the LAP (see \hyperref[methods]{Methods}) with an effective object-space resolution of 40\,\um\,/\,px. After three weeks, much more regions show diattenuation of type $D^+$ (green) and the main peak of ($\phiD - \phiP$) is shifted from about $90^{\circ}$ to $0^{\circ}$.}
	\label{fig:Diattenuation_vs_Time_Human}
\end{figure}


\begin{figure}[H]
	\centering
	\includegraphics[width=\textwidth]{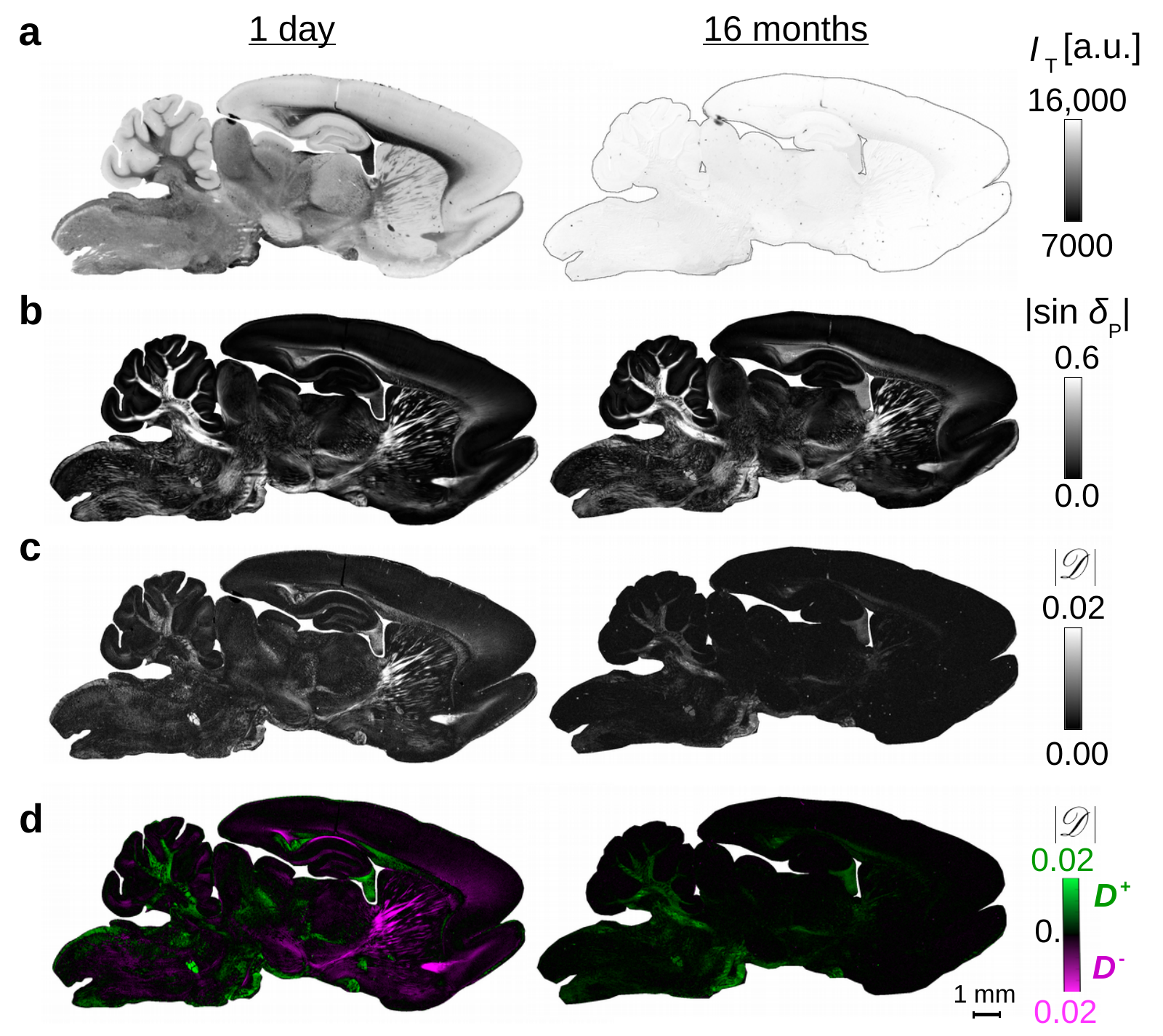}
	\caption{Dependence of transmittance, birefringence, and diattenuation on embedding time (sagittal rat brain section in Figs.\ \ref{fig:DI}c and \ref{fig:DiattenuationMaps_Coronal-Sagittal}a measured one day and 16 months after embedding). \textbf{a} Average transmitted light intensity (transmittance $I_{\text{T}}$). \textbf{b} Strength of the measured birefringence signal $|\sin\delta_{\text{P}}|$. \textbf{c} Strength of the measured diattenuation signal $\Dm$. \textbf{d} Diattenuation images: $\Dm$ values belonging to regions with diattenuation of type $D^{+}$ ($\phiD - \phiP \in [-20^{\circ},20^{\circ}]$) are shown in green, values belonging to regions with diattenuation of type $D^{-}$ ($\phiD - \phiP \in [70^{\circ},110^{\circ}]$) are shown in magenta). The values $\{I_{\text{T}}, |\sin\delta_{\text{P}}|, \Dm, \phiD, \phiP\}$ were determined from polarimetric measurements with the LAP (see \hyperref[methods]{Methods}) with an effective object-space resolution of 14\,\um\,/\,px. After 16 months, the transmittance fades out and regions with $D^-$ vanish, while the birefringence barely changes.}
	\label{fig:Diattenuation_vs_Time_Rat}
\end{figure}


\begin{figure}[H]
	\centering
	\includegraphics[width=0.9\textwidth]{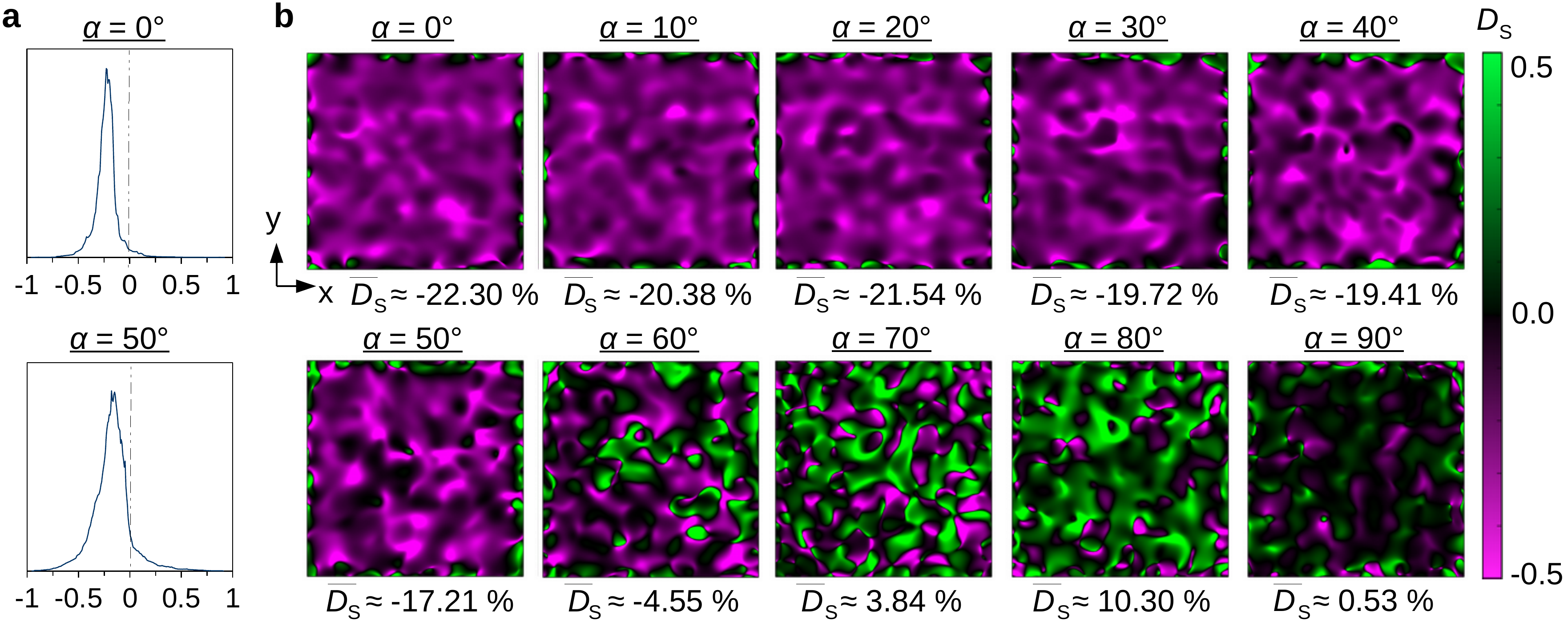}
	\caption{Simulated diattenuation images for different fibre inclination angles (for the bundle of densely grown fibres, see \cref{fig:Diattenuation_Sim_vs_Exp}a). \textbf{a} Histograms of $\DS$ for inclination $\alpha=0^{\circ}$ (mean: $-22\,\% \pm 12\,\%$) and $\alpha = 50^{\circ}$ (mean: $-17\,\% \pm 19\,\%$). \textbf{b} Diattenuation images $\DS$ and mean values $\overline{\DS}$ for all inclination angles. Positive diattenuation values (representing type $D^+$) are displayed in green, negative diattenuation values (representing type $D^-$) are displayed in magenta. The diattenuation images were obtained from simulated diattenuation measurements with light polarised along the x-axis ($I_{\text{x}}$) and along the y-axis ($I_{\text{y}}$): $\DS = (I_{\text{x}} - I_{\text{y}})/(I_{\text{x}} + I_{\text{y}})$ (see \hyperref[methods]{Methods}). The diattenuation is mostly negative (magenta) for fibres with small inclination angles ($\alpha \leq 50^{\circ}$) and becomes more positive (green) for steeper fibres ($\alpha > 60^{\circ}$).}
	\label{fig:Sim_DiattenuationMaps}
\end{figure}


\newpage
\subsection*{Supplementary Note 1. Analytical Model of Dichroism}
\label{note1}


\subsubsection*{Dichroism in Uniaxial Absorbing Materials}

In uniaxial absorbing materials with high symmetry, the birefringence (anisotropic refraction) and the dichroism (anisotropic absorption) can be described by a complex retardance with shared principal axes.

Defining a complex refractive index
\begin{align}
n' \equiv n + \I \kappa \equiv n ( 1 + \I \hat{\kappa})
\label{eq:n}
\end{align}
and assuming weak absorption ($\kappa \ll n$), the absorption of the extraordinary light wave ($\hat{\kappa}_{\text{e}}$) is given by (cf.\ \textsc{Born} \& \textsc{Wolf} \cite{born}, Sec.\ 15.6.1):
\begin{align}
v_{\text{e}}^2 \, \hat{\kappa}_{\text{e}} &= \vo^2 \, \hat{\kappa}_{\text{o}} \, \cos^2\theta + \vE^2 \, \hat{\kappa}_{\text{E}} \, \sin^2\theta,
\label{eq:Appendix_kappa2}
\end{align}
where $\theta$ is the angle between the optic axis (symmetry axis) and the wave vector, $v$ is the phase velocity of the light, and the indices ``o'', ``e'', and ``E'' denote the properties of the ordinary wave (polarised perpendicularly to the optic axis), the extraordinary wave, and the principal extraordinary wave (polarised parallel to the optic axis), respectively.

Writing $\hat{\kappa} = \kappa / n$, $v=c/n$ ($c$ is the velocity of light in vacuum), and defining $\dn \equiv \nE - \no$, $\dn_{\text{e}} \equiv \nex - \no$, $\dk \equiv \kE - \ko$, \cref{eq:Appendix_kappa2} yields:
\enlargethispage{0.8cm}
\begin{align}
\kex &= \frac{\nex^3}{\no^3} \, \ko \, \cos^2\theta + \frac{\nex^3}{\nE^3} \, \kE \, \sin^2\theta \\
&= \underbrace{\frac{(\no + \dn_{\text{e}})^3}{\no^3}}_{f_1\left(\dn_{\text{e}}\right)} \, \ko \, \cos^2\theta + \underbrace{\frac{(\no + \dn_{\text{e}})^3}{(\no + \dn)^3}}_{f_2\left(\dn_{\text{e}},\,\dn\right)} \, (\ko + \dk) \, \sin^2\theta. 
\label{eq:Appendix_kappa3}
\end{align}

The birefringence of biological tissue \cite{ghosh2011} ($|\dn| = 0.001$--$0.01$) is small compared to its refractive index values \cite{beuthan1996} ($n=1.3$--$1.5$). 
In this case, a first order Taylor expansion (in 2D) can be applied to the functions $f_1$ and $f_2$ in \{$\dn_{\text{e}}=0$, $\dn=0$\}:
\begin{align}
f_1\left(\dn_{\text{e}}\right) &= f(0) + f'(0) \, \dn_{\text{e}} + \dots \\
&= 1 + \frac{3}{\no}\,\dn_{\text{e}} + \dots 
\label{eq:Appendix_f1} \\
f_2\left(\dn_{\text{e}},\,\dn\right) &= f(0,0) + \frac{\partial}{\partial \dn_{\text{e}}} \, f(0,0) \, \dn_{\text{e}} + \frac{\partial}{\partial\dn} \, f(0,0) \, \dn + \dots \\
&= 1 + \frac{3}{\no} \left( \dn_{\text{e}} - \dn \right) + \dots 
\label{eq:Appendix_f2}
\end{align}
For $|\dn| \ll n$, the following approximation can be used, see \textsc{Menzel} \ea\ \cite{menzel2015}, equation A5:
\begin{align}
\dn_{\text{e}} \approx \dn \, \sin^2\theta.
\label{eq:birefringence}
\end{align}

Inserting \cref{eq:Appendix_f1,eq:Appendix_f2} into \cref{eq:Appendix_kappa3} and using \cref{eq:birefringence} yields:
\begin{align}
\kex &\approx \left( 1 + \frac{3}{\no} \dn \, \sin^2\theta \right) \ko \cos^2\theta
+ \left( 1 - \frac{3}{\no} \dn \, \cos^2\theta \right)(\ko + \dk)\sin^2\theta \\
&= \ko + \left( 1 - \frac{3}{\no} \dn \, \cos^2\theta \right) \dk \, \sin^2\theta.
\label{eq:kappa_approx}
\end{align}
For $|\dn| \leq 0.01$ and $n \geq 1.3$, the term in round brackets in \cref{eq:kappa_approx} is $> 0.97$ and $\kex \approx \ko + \dk \, \sin^2\theta$.

Thus, in uniaxial absorbing materials with small birefringence ($|\dn| \ll n$) and weak absorption ($\kappa \ll n$) like brain tissue, the dichroism (anisotropic absorption) depends on the angle $\theta$ between the wave vector and the optic axis, just like the birefringence in \cref{eq:birefringence}:
\begin{align}
\dk_{\text{e}} \equiv \kex - \ko \approx \dk \, \sin^2\theta \, , 
\quad \dk \equiv \kE - \ko.
\label{eq:Appendix_dk}
\end{align}


\newpage
\subsubsection*{Inclination Dependence of Dichroism}
The diattenuation of a brain section is computed via:
\begin{align}
D = \frac{I_{\parallel} - I_{\perp}}{I_{\parallel} + I_{\perp}},
\label{eq:D_}
\end{align}
with $I_{\parallel}$ and $I_{\perp}$ being the (maximum and minimum) transmitted light intensities for light polarised parallel and perpendicularly to the projection of the optic axis (predominant orientation of the nerve fibres) onto the section plane.

For light propagating through a brain section of thickness $d$, the transmitted light intensity is given by the Beer-Lambert law:
\begin{align}
I &= I_0 \, \e^{- \mu \, d}, 
\end{align}
where $I_0$ is the ingoing light intensity and $\mu$ the attenuation coefficient of the material which depends on the direction of polarisation.
With this definition, \cref{eq:D_} can be written as:
\begin{align}
D &= \frac{\e^{-d\,\mu_{\parallel}} - \e^{-d\,\mu_{\perp}}}{\e^{-d\,\mu_{\parallel}} + \e^{-d\,\mu_{\perp}}}
= \frac{\e^{\,d\,(\mu_{\perp} - \mu_{\parallel})/2} - \e^{-d\,(\mu_{\perp} - \mu_{\parallel})/2}}{\e^{\,d\,(\mu_{\perp} - \mu_{\parallel})/2} + \e^{-d\,(\mu_{\perp} - \mu_{\parallel})/2}} \\
&= \tanh \left( d \,\, \frac{\mu_{\perp} - \mu_{\parallel}}{2} \right).
\label{eq:D_tanh}
\end{align}

To estimate how the dichroism of brain tissue depends on the nerve fibre inclination, we assume that the diattenuation is solely caused by anisotropic absorption, neglecting any scattering.
In this case, the attenuation coefficient is given by the absorption coefficient (\textsc{Demtröder} \cite{demtroeder}, p.\ 227):
\begin{align}
\mu = K \equiv \frac{4 \pi \kappa}{\lambda},
\label{eq:abs}
\end{align}
where $\kappa$ is the imaginary part of the refractive index in the medium (see \cref{eq:n}) and $\lambda$ the wavelength of the light.
In a similar way, we define the absorption coefficient of the ordinary wave ($K_{\perp} \equiv 4 \pi \ko / \lambda$) and of the extraordinary wave ($K_{\parallel} \equiv 4 \pi \kex / \lambda$), taking into account that the ordinary wave is polarised perpendicularly to the optic axis while the extraordinary wave is polarised parallel to the projection of the optic axis onto the section plane. 

With these definitions, the diattenuation caused by anisotropic absorption (dichroism) can be written as:
\begin{align}
\DK \overset{(\ref{eq:D_tanh})}{=} \tanh \left( \frac{d}{2} \big(K_{\perp} - K_{\parallel}\big) \right) 
\overset{(\ref{eq:abs})}{=} \tanh \left(\frac{2\pi d}{\lambda} \big(\ko - \kex\big) \right) 
\overset{(\ref{eq:Appendix_dk})}{\approx} \tanh \left(-\frac{2\pi d}{\lambda} \, \dk \, \sin^2\theta\right),
\label{eq:dichroism}
\end{align}
where $\theta$ is the angle between the optic axis (nerve fibre orientation) and the wave vector (direction of propagation).

Assuming that the brain section is illuminated under normal incidence, the out-of-plane inclination angle of the fibres is given by: $\alpha = 90^{\circ} - \theta$. Thus, the dichroism of brain tissue decreases with increasing fibre inclination angle and does not change its sign (cf.\ \cref{fig:Diattenuation_Sim_vs_Exp}c):
\begin{align}
\DK \approx \tanh \left(-\frac{2\pi d}{\lambda} \, \dk \, \cos^2\alpha \right).
\label{eq:DK}
\end{align}

Freshly embedded brain sections show diattenuation of both types $D^+$ and $D^-$ (see \cref{fig:DiattenuationMaps_Coronal-Sagittal}). This suggests that the diattenuation is not only caused by dichroism (anisotropic absorption) but also by anisotropic scattering of light.
With increasing time after tissue embedding, the brain sections become transparent (see Supplementary \cref{fig:Diattenuation_vs_Time_Rat}a), \ie the scattering is expected to decrease due to an equalisation of the refractive indices. Thus, the diattenuation of type $D^+$ that is observed in brain sections with long embedding time (see \cref{fig:Diattenuation_vs_Time_Vervet} and Supplementary \cref{fig:Diattenuation_vs_Time_Human}) is presumably caused by dichroism, \ie $\DK > 0 \Leftrightarrow I_{\parallel} > I_{\perp}$. This means that the absorption becomes maximal (the transmitted light intensity becomes minimal) when the light is polarised perpendicularly to the fibre axis, \ie in the plane of the radially oriented lipid molecules in the myelin sheaths. This suggests that dichroism of brain tissue is mainly caused by the myelin lipids and is therefore not expected to change with increasing time after tissue embedding, just like the birefringence (see Supplementary \cref{fig:Diattenuation_vs_Time_Rat}b).


\end{document}